\documentclass[12pt]{article}
\usepackage{amsmath, amssymb, graphicx,epsfig}
\usepackage{verbatim}
\usepackage{bbm}
\usepackage{appendix}
\usepackage{xcolor}
\usepackage{feynmp}

\usepackage{amsmath}
\usepackage{amssymb}
\usepackage{graphicx}
\usepackage{verbatim}
\usepackage{cancel}
\usepackage{bbm}
\usepackage{xcolor}
\usepackage{feynmp}

\usepackage{float}
\usepackage{color}
\restylefloat{table}

\def\id{\protect{{1 \kern-.28em {\rm l}}}}

\makeatletter
\renewcommand\section{\@startsection {section}{1}{\z@}%
                                   {-3.5ex \@plus -1ex \@minus -.2ex}%
                                   {2.3ex \@plus.2ex}%
                                   {\normalfont\large\bfseries}}
\renewcommand\subsection{\@startsection{subsection}{2}{\z@}%
                                   {-3.25ex\@plus -1ex \@minus -.2ex}%
                                   {1.5ex \@plus .2ex}%
                                   {\normalfont\normalsize\bfseries}}
\makeatother

\newcommand{\bea}{\begin{eqnarray}}
\newcommand{\eea}{\end{eqnarray}}

\def\be{\begin{eqnarray}}
\def\ee{\end{eqnarray}}
\def\la{\label}

\def\k{\varkappa}
\def\L{{\cal L}}
\def\ov{\over}

\def\a {\alpha}
\def\b {\beta}

\def\g {\gamma}

\def\p{\phi}

\def\z{\zeta}
\def\eps{\epsilon}
\def\r {\rho}

\def\pa {\partial}

\def\STr{{\rm STr}}

\def\cT{{\mathcal T}}
\def\de{\delta }

\newcommand{\alg}[1]{\mathfrak{#1}}

\newcommand{\smatrix}{\mathbf{S}}   
\newcommand{\tmatrix}{\cT}   

\newcommand{\unit}{\mathbbm{1}}
\newcommand{\order}{\mathcal{O}}

\begin{document}


\overfullrule=0pt
\parskip=2pt
\parindent=12pt
\headheight=0in \headsep=0in \topmargin=0in \oddsidemargin=0in

\vspace{ -3cm}
\thispagestyle{empty}
\vspace{-1cm}

\rightline{UUITP-21/14}
\rightline{NSF-KITP-14-207}

\

\

\begin{center}
\vspace{1cm}
{\Large\bf  
On the asymptotic states and the quantum S matrix of the $\eta$-deformed $AdS_5\times S^5$ superstring
}

\end{center}

\vspace{.2cm}



\begin{center}
 
Oluf Tang Engelund$^a$ and   Radu Roiban$^{b,c}$

\end{center}

\begin{center}
{
\em 
${}^a$ Department of Physics and Astronomy, Uppsala University,\\
 SE-751 08 Uppsala, Sweden\\
 {\tt oluf.engelund@physics.uu.se}
\vskip 0.08cm
\vskip 0.08cm
${}^b$ Department of Physics, The Pennsylvania  State University,\\
University Park, PA 16802 , USA\\
{\tt radu@phys.psu.edu}
\vskip 0.08cm
\vskip 0.08cm 
${}^c$ Kavli Institute for Theoretical Physics, University of California\\
Santa Barbara, CA 93106-4030 USA
}
 \end{center}



\vspace{1.5cm}

\vspace{.2cm}

\begin{abstract}

\noindent 

We investigate the worldsheet S matrix of string theory in $\eta$-deformed $AdS_5\times S^5$. 
By computing the six-point tree-level S matrix we explicitly show that there is no particle production 
at this level, as required by the classical integrability of the theory. At one  and two loops we show that 
integrability requires that the classical two-particle states be redefined in a non-local and 
$\eta$-dependent way. This is a significant departure from the undeformed theory which is probably 
related to the quantum group symmetry of the worldsheet theory. 
We use generalized unitarity to carry out the loop calculations and identify a set of integrals that
allow us to give a two-loop Feynman integral representation of the logarithmic terms of the two-loop 
S matrix. We finally also discuss aspects of the calculation of the two-loop rational terms.

\end{abstract}

\newpage



\section{Introduction}

Integrability of the string sigma model is a key feature that makes possible the determination of the string spectrum 
on non-trivial curved backgrounds \cite{beir}. It is therefore important to identify and analyze such sigma models which correspond to
physically-interesting string theories. Examples are integrable deformations of string sigma models on 
AdS$_n\times$S$^n\times$M$^{10-2n}$ which, in the undeformed case, play an important role  in the AdS/CFT 
correspondence. 

Orbifolding or sequences  of T-duality (or worldsheet duality) and shift transformations (see {\it e.g.} \cite{lm, frt, fa, rtw, bei}) of 
an integrable two-dimensional sigma model provide a straightforward way of constructing closely related integrable models.
Generalizing previously-known constructions of integrable deformations of group or coset models \cite{dmv0, k1, k2, foz, fat, luk},
a classically-integrable deformation of the AdS$_5\times$S$^5$ Green-Schwarz sigma model was proposed in \cite{Delduc:2013qra}.
%
%
The deformation completely breaks target space supersymmetry and reduces the AdS$_5$ and S$^5$ isometries to their 
Cartan subgroups, U$(1)^3\otimes $U$(1)^3$. Remarkably however, the original symmetry is not completely lost but rather 
it is $q$-deformed to PSU$_q(2,2|4)$ \cite{Delduc:2014kha}.

The bosonic Lagrangian was constructed explicitly and it was quantized in uniform light-cone gauge 
in ref.~\cite{Arutyunov:2013ega} (see  \cite{Hoare:2014pna} for lower-dimensional models and \cite{Lunin:2014tsa} for a 
discussion  of the corresponding supergravity backgrounds); the bosonic tree-level S matrix was also constructed and 
%
%
shown to reproduce  
the small momentum (classical) limit of the PSU$_q(2|2)^2$-symmetric S matrix of  \cite{Beisert:2008tw, beis, hhmphase}, suggesting 
that the gauge-fixed theory has indeed this symmetry. 
Integrability of the theory implies then that, if this symmetry is preserved at the quantum level, the S matrix should factorize as \cite{Zamolodchikov:1978xm}
\be
\mathbb{S}=\smatrix_{\text{PSU}_q(2|2)}\otimes \smatrix_{\text{PSU}_q(2|2)} \ ,
\ee
where each factor is invariant under a different PSU$_q(2|2)$ factor and may be written as
\begin{align}
\label{SmatExpansion}
\smatrix_{\text{PSU}_q(2|2)} =& e^{i{\hat \theta}_{12}}{\hat \smatrix}_{\text{PSU}_q(2|2)}  
\equiv \unit+\frac{i}{g}\tmatrix 
=e^{i{\hat \theta}_{12}} \left(\unit+\frac{i}{g} \hat\tmatrix  \right)
\\
=&\unit+\frac{1}{{g}}i{\tmatrix}^{(0)}+\frac{1}{{g}^2}i\left({\hat\tmatrix}^{(1)}+\frac{1}{2}{\hat\theta}^{(1)}_{12}\unit\right)
+\frac{1}{{g}^3}i\left({\hat\tmatrix}^{(2)}+\frac{i}{2} {{\hat\theta}^{(1)}}_{12}{\tmatrix}^{(0)}
+\frac{1}{2}{{\hat\theta}^{(2)}}_{12}\unit \right)+ \order\left({\frac{1}{{g}^4} }\right) \ .
\nonumber
\end{align}
Here ${\hat \smatrix}_{\text{PSU}_q(2|2)}$ is the part of the S matrix determined by the symmetries normalized such that the
dressing phase is unity at tree level.
 
The small amount of manifest symmetry in this theory suggests that, by studying it, we may expose features that did not appear 
in the undeformed theory. For example, it is interesting to wonder whether integrability survives at higher orders and how is the 
PSU$_q(2|2)^2$ realized at the quantum level on the Lagrangian fields. The perturbative worldsheet S matrix is perhaps the 
most basic quantity which may help address these questions.
%
%
We will compute it  at tree-level beyond leading nontrivial S-matrix elements, as well as at one- and two-loop order. 
In doing so we shall also identify an integral basis which, in conjunction with generalized unitarity,  yields a Feynman 
integral representation for all the logarithmic terms in the two-loop S matrix. The construction of this basis may be iterated 
to all loop orders.

An important property of higher-point S matrices in integrable theories is the absence of particle production or, alternatively, 
their factorization of the (tree-level) higher-point S matrix into sequences of $2\rightarrow 2$ processes \cite{Zamolodchikov:1978xm}.
This feature has important simplifying consequences on the unitarity-based construction of the S matrices of such 
theories \cite{Bianchi:2013nra, Engelund:2013fja, Bianchi:2014rfa}. As we shall review in sec.~\ref{finiteness}, it implies the 
cancellation of massive tadpole integral contributions to the 1PI part of the S matrix and thus suggests that, if present, UV divergences
are confined to the renormalization of two-point functions. 

%
It was pointed out  in \cite{Arutyunov:2006yd} that, for an S matrix to have desirable properties, one should in principle allow for 
transformations of the multi-particle scattering basis, which from the point of view of the constituent one-particle states appears 
mutually non-local. These transformations may significantly modify the symmetry properties of the S matrix without changing 
the actual physical content.
As we shall see, such a bilocal transformation (in momentum space) is necessary in the $\eta$-deformed theory to put the 
loop-level S matrix in the form \eqref{SmatExpansion} suggested by the integrability and classical symmetries of the theory. 
One may, alternatively, interpret the required transformation as acting on single-particle states at the expense of changing 
their dimension and spin, both of which become formally complex. 
The necessity for this redefinition is a significant departure from the undeformed theory\footnote{In the undeformed theory a redefinition 
of creation/annihilation operators is necessary to relate the worldsheet and spin chain S matrices, see \cite{Arutyunov:2006yd}.} 
and appears to be closely related to 
the presence of an NS-NS $B$ field and the corresponding bosonic Wess-Zumino term. 
However, the presence of such a field does not necessarily require such  redefinition as shown by 
loop calculations in AdS$_3\times$S$^3\times$T$^4$ supported by mixed flux \cite{Hoare:2013pma, Hoare:2013ida}. It therefore 
seems likely that it is required for 
the naive tree-level asymptotic states to become a representation of ${\text{PSU}_q(2|2)}$.

In general, to carry out loop calculations it is necessary to know the interaction terms containing worldsheet fermions. As we 
shall see however, part of our conclusions can be reached based only on the structure of the S matrix and with minimal detailed 
information on the fermion-dependent part of the Lagrangian or of the corresponding S-matrix elements. When we derive explicit 
expressions of loop-level S matrix we shall use for the currently-unknown tree-level S matrix the relevant terms in the small 
momentum expansion of the ${\text{PSU}_q(2|2)}$ S matrix of~\cite{Beisert:2008tw, beis, hhmphase}.

The paper is organized as follows. In sec.~\ref{Lagrangian} we review the deformed Lagrangian and its bosonic part, the structure of 
the four-particle S matrix and discuss the factorization of the six-particle S matrix. In sec.~\ref{oneloop} we construct the  one-loop S matrix
in terms of the tree-level S-matrix coefficients and identify the redefinition of the two-particle states that cast it in the form suggested by 
the classical symmetries and integrability. In sec.~\ref{twoloop} we describe a new basis of two-loop integrals, give an integral representation
of the logarithmic terms of the two-loop S matrix and provide a discussion of the rational terms.
In sec.~\ref{Diskussion} we summarize our results and discuss how to construct an integral representation for the worldsheet S matrix 
at arbitrary loop order.
We relegate to appendices explicit expressions for the tree-level S-matrix coefficients, one-loop integral 
coefficients and one-loop S-matrix coefficients and explicit expressions for one- and two-loop integrals.

\section{The deformed action and bosonic Lagrangian \label{Lagrangian}}

The one-parameter $\eta$-deformation of  the AdS$_5\times$S$^5$ supercoset Lagrangian  
constructed in  \cite{Delduc:2013qra} is naturally expressed in terms of the left-invariant 
one-forms of the undeformed symmetry group:
\be
\label{de}
&& 
{ L} =  c_\eta\,  \pi^{ij}\, \STr[J_i \, d_\eta \circ \frac{1}{1-\eta R_g\circ d_\eta}\, J_j] 
\qquad\qquad
 \pi^{ij} \equiv \sqrt {-h} h^{ij} - \epsilon^{ij} 
\ , \\
&&J_i =g^{-1}\partial_i g
 \qquad \qquad
d_\eta \equiv P_1+ 2 c_\eta^{-1} 
P_2 -P_3\ ,
~~~~\qquad
c_\eta \equiv  1-\eta^2 \ .
\label{def}
\ee
Here $g\!\in\,$PSU$(2,2|4)$ and $P_k$ are projectors onto subspaces with eigenvalue $i^k$ of the action of the $Z_4$ automorphism 
of PSU$(2,2|4)$.\footnote{We use the normalization in which the (super)trace of squares of the bosonic
Cartan generators equals 2.}
The operator $R_g$ acts on the superalgebra as
\be
R_g(M) = g^{-1}R(g M g^{-1})g \ ,
\label{Rgaction}
\ee
where the operator $R$ multiplies the generators
corresponding to the positive roots by $-i$, those corresponding to the negative roots by $+i$
and annihilates the Cartan generators.
There are three choices of $R$ operator leading to inequivalent bosonic actions (the 
corresponding metrics appear to have different singularity structures) \cite{Delduc:2014kha}.

The Lagrangian \eqref{def} has several remarkable properties. On the one hand it 
preserves the classical integrability of the undeformed theory. On the other, it exhibits a $q$-deformed 
symmetry \cite{Delduc:2014kha}, which suggests that  the theory is more symmetric than 
manifest from the Lagrangian. The parameter $q$ is related to the deformation parameter 
$\eta$ as
\be
q = e^{-\nu/g}
\qquad
\nu = \frac{2\eta}{1+\eta^2} \ .
\ee
This relation was initially inferred in \cite{Arutyunov:2013ega} by comparing the tree-level S matrix of the 
deformed model with the PSU$_q(2|2)^2$-invariant S matrix of \cite{Beisert:2008tw, beis, hhmphase}. 
Up to the normalization of the worldsheet action (and hence of $g$), the same expression was found in 
 \cite{Delduc:2014kha} where the symmetries of the classical action have been analyzed.

\subsection{The bosonic Lagrangian and the four-point S matrix}

Using the choice of $R$ operator put forth in \cite{Delduc:2013qra} and a judicious parameterization of the coset,  
the bosonic Lagrangian was constructed in \cite{Arutyunov:2013ega}. Unlike the undeformed theory, the geometric 
background is supplemented by a nontrivial NSNS B-field. 
The Lagrangian is:
\be
\mathcal{L}=\L_{\alg{a}}^{G} +\L_{\alg{s}}^{G} +\L_{\alg{a}}^{WZ} +\L_{\alg{s}}^{WZ} 
\ee
with\footnote{The relation between $\k $ and $\eta$ is $\eta = \k ^{-1} [\sqrt{ 1 + \k ^{2}} - 1 ]$.} 
\bea\nonumber
\L_{\alg{a}}^{G} &=&-{g\ov2}(1+\varkappa^2)^{1\ov2}\, \g^{\a\b}\Big(
-\frac{\pa_\a t\pa_\b t\left(1+\rho ^2\right)}{ 1-\varkappa ^2 \rho ^2}
+\frac{\pa_\a \r\pa_\b \r}{ \left(1+\rho ^2\right) \left(1-\varkappa ^2 \rho ^2\right)}
+\frac{\pa_\a \z\pa_\b\z \rho ^2}{1+ \varkappa ^2 \rho ^4 \sin ^2\z }
\\\la{LaG}
   &&\qquad\qquad\qquad+\frac{\pa_\a \psi_1\pa_\b\psi_1\rho ^2 \cos
   ^2\z}{ 1+\varkappa ^2 \rho ^4 \sin ^2\z}+\pa_\a \psi_2\pa_\b\psi_2
  \rho ^2 \sin ^2\z \Big)\,,
\\[2pt]
\nonumber
\L_{\alg{s}}^{G} &=&-{g\ov2}(1+\varkappa^2)^{1\ov2}\, \g^{\a\b}\Big(\frac{\pa_\a \p\pa_\b \p
  \left(1-r^2\right)}{1+\varkappa ^2 r^2}+\frac{\pa_\a r\pa_\b r
  }{ \left(1-r^2\right) \left(1+\varkappa ^2 r^2\right)}
  +\frac{\pa_\a \xi\pa_\b \xi  r^2}{1+ \varkappa ^2 r^4 \sin ^2\xi}
  \\\la{LsG}
   &&\qquad\qquad\qquad+\frac{\pa_\a \p_1\pa_\b \p_1  r^2 \cos ^2\xi }{1+ \varkappa ^2
   r^4 \sin ^2\xi } +\pa_\a \p_2\pa_\b \p_2  r^2 \sin^2\xi \Big)\,,
\eea
and the Wess-Zumino terms $\L_{\alg{a}}^{WZ}$ and $\L_{\alg{s}}^{WZ}$ given by
\bea
\la{LaWZ}
\L_{\alg{a}}^{WZ} &=&{g\ov2} \varkappa (1+\varkappa^2)^{1\ov2}\, \eps^{\a\b}\frac{ \rho ^4 \sin 2 \zeta}{1+ \varkappa ^2 \rho ^4 \sin ^2\z}\pa_\a\psi_1\pa_\b\zeta\,,
\\[2pt]
\la{LsWZ}
\L_{\alg{s}}^{WZ} &=&-{g\ov2} \varkappa (1+\varkappa^2)^{1\ov2}\, \eps^{\a\b}\frac{ r^4 \sin 2 \xi }{1+ \varkappa ^2 r^4 \sin^2\xi}\pa_\a\p_1\pa_\b\xi\,  .
\eea

The light-cone gauge-fixing of this Lagrangian was discussed at length in \cite{Arutyunov:2013ega} and we will not reproduce it here. 
For the purpose of the construction of the S matrix it is useful to pass to complex coordinates, which manifest the SU$(2)^4$ in the 
$\kappa\rightarrow 0$ limit. Restricting to the $S^5$ fields the transformation is
\bea
&&
r = \frac{|y|}{1+\frac{1}{4}y^2 }
~~,\qquad
\cos^2\xi = \frac{y_1^2+y_2^2}{y^2}
~~,\qquad
\sin^2\xi = \frac{y_3^2+y_4^2}{y^2}
\\
&&
Y^{1\dot{1}}={}\tfrac{1}{2}\left(y_1+iy_2\right) ~,\quad Y^{2\dot{2}}={}\tfrac{1}{2}\left(y_1-iy_2\right)
~,\quad 
Y^{1\dot{2}}={}\tfrac{1}{2}\left(y_3-iy_4\right)  ~,\quad Y^{2\dot{1}}={}-\tfrac{1}{2}\left(y_3+iy_4\right)
\nonumber \ .
\eea
The Lagrangian to quadratic and quartic orders ($Y^2 = 4(Y^{1{\dot 1}}Y^{2{\dot 2}}-Y^{2{\dot 1}}Y^{1{\dot 2}})$, etc) is then\footnote{These expressions 
are obtained by Legendre-transforming the Hamiltonian of \cite{Arutyunov:2013ega}. Alternative expressions may be obtained by expanding the 
Nambu-Goto action.}
\begin{align}
\mathcal{L_S}=&\mathcal{L}_{2,S}+\mathcal{L}_{4,S}+\mathcal{L}_{4,S}^{WZ}+\dots
\\
\mathcal{L}_{2,S} =& \frac{1}{2}g\left(-\partial_0 Y^{\alpha\dot\beta}\partial_0 Y_{\alpha\dot\beta} 
+(1+\varkappa^2) \partial_1 Y^{\alpha\dot\beta}\partial_1 Y_{\alpha\dot\beta}+ (1+\varkappa^2) Y^{\alpha\dot\beta}Y_{\alpha\dot\beta} \right)
\label{action_quadratic}
\\
\mathcal{L}_{4,S} =&-\frac{1}{2}g(1+\varkappa^2) Y^2 (\partial_1 Y)^2+\frac{1}{2}g\varkappa^2Y^2 (\partial_0 Y)^2
%
\label{quarticeven}
\\
\mathcal{L}_{4,S}^{WZ}=&{}2ig\varkappa\sqrt{1+\varkappa^2}\; Y^{1\dot{2}}Y^{2\dot{1}}\epsilon^{\alpha\beta}\big(\partial_\alpha Y^{1\dot{1}}\big)\big(\partial_\beta Y^{2\dot{2}}\big) \ .
\label{quarticodd}
\end{align}
Remarkably, the bosonic tree-level four-point S matrix given by this Lagrangian reproduces  \cite{Arutyunov:2013ega} 
the small momentum limit of the exact PSU$_q(2|2)^2$-invariant S matrix of \cite{Beisert:2008tw, beis, hhmphase}.

In secs.~\ref{oneloop} and \ref{twoloop} we shall need the general form of the two-particle S matrix. 
Based on the manifest and expected symmetries the general form of the $\tmatrix$-matrix elements in \eqref{SmatExpansion} is:
\bea\la{cTmatr}
\begin{aligned}
&\cT{}_{ab}^{cd}=
A\,\de_a^c\de_b^d+\de_a^d\de_b^c(B+W_B\, \eps_{ab} - V_B \epsilon_{ab}\epsilon^{cd})\, ,   \\ 
&\cT{}_{\a\b}^{\g\de}=
D\,\de_\a^\g\de_\b^\de+\de_\a^\de\de_\b^\g(E+W_E\, \eps_{\a\b}- V_E \epsilon_{\alpha\beta}\epsilon^{\gamma\delta})\,,  \\
&\cT{}_{ab}^{\g\de}=
\epsilon_{ab}\epsilon^{\g\de}(C+Q_C\epsilon_{ab} -Q_C \epsilon^{\g\de}+R_C\epsilon_{ab}\epsilon^{\g\de})\,, \\
&\cT{}_{\a\b}^{cd}=
\epsilon_{\a\b}\epsilon^{cd}(F+Q_F\epsilon_{\a\b} -Q_F \epsilon^{cd}+R_F\epsilon_{\a\b}\epsilon^{cd})\,, \\
&\cT{}_{a\b}^{c\de}= G\,\de_a^c\de_\b^\de \  ,\qquad
~\cT{}_{\a b}^{\g d}= L\,\de_\a^\g\de_b^d \ ,\\
&\cT{}_{a\b}^{\g d}= H\,\de_a^d\de_\b^\g \  ,\qquad
~\cT{}_{\a b}^{c \de}= K\,\de_\a^\de\de_b^c \ .
\end{aligned}
\eea 
The tree-level values of the coefficients of the bosonic structures, $A, B, D, E, G, L, W$, have been constructed directly 
from the Lagrangian in \cite{Arutyunov:2013ega}.  At this order
\be
W^{(0)}_B={}W^{(0)}_E =W^{(0)}=i\nu\ ;
\label{Wtree}
\ee
their common value $W^{(0)}$ corresponds to the contribution of the Wess-Zumino  term and it does not depend on the particle momenta.
%
%
In Appendix~\ref{TreeSmatrix} we collect the tree-level expressions of all coefficients in \eqref{cTmatr} 
extracted from  \cite{Beisert:2008tw, beis, hhmphase} by taking the small momentum expansion.

\subsection{Six-point S matrix and absence of particle production}

One of the consequences of integrability is the absence of particle production or, alternatively, the factorization of the 
$n\rightarrow n$ S matrix into a sequence of $2\rightarrow 2$ scattering events \cite{Zamolodchikov:1978xm}; 
all possible factorizations are equivalent as a consequence of the Yang-Baxter equation obeyed by the four-particle 
S matrix. 
Here we discuss the absence of $2 \rightarrow 4$ tree-level scattering processes for the $\eta$-deformed 
worldsheet theory and the corresponding factorization of the $3 \rightarrow 3$ tree-level amplitude.
This calculation verifies the classical integrability of the gauge-fixed theory and, moreover, is an integral part of
the unitarity-based approach to the construction of the S matrix in integrable quantum field theories.

For the purpose of illustration we will focus here on the fields parametrizing S$^5$.
It is straightforward, albeit tedious, to expand the parity-even part of the gauge-fixed deformed Lagrangian to this order. 
It is however simplest to check the factorization of the 
parity-odd part of the (bosonic) S matrix. Indeed, these matrix elements depend only of the parity-odd six-field terms in the 
expansion of the Lagrangian (and lower order terms as well) which are substantially simpler. In the notation of  
\cite{Arutyunov:2013ega}, they are given by:
\begin{align}
\mathcal{L}_s^{WZ}={}48ig\varkappa(1-\varkappa^2)\sqrt{1+\varkappa^2}\left(Y^{1\dot{2}}Y^{2\dot{1}} -2Y^{1\dot{1}}Y^{2\dot{2}} \right)
Y^{1\dot{2}}Y^{2\dot{1}}\epsilon^{\alpha\beta}\big(\partial_\alpha Y^{1\dot{1}}\big)\big(\partial_\beta Y^{2\dot{2}}\big)+\mathcal{O}(X^8) \ .
\label{WZ6pt}
\end{align}
The propagator coming from the quadratic Lagrangian is of the form
\bea
\pm i\Delta=\frac{\pm i}{\omega_q^2-\alpha q^2-m^2}.
\eea
for some choice of $\alpha$ and $m$. The Feynman rules from the quartic Lagrangian \eqref{quarticeven}-\eqref{quarticodd} are
\begin{align}
\hbox{\raise-18pt\hbox{~\mbox{\includegraphics[height=17mm]{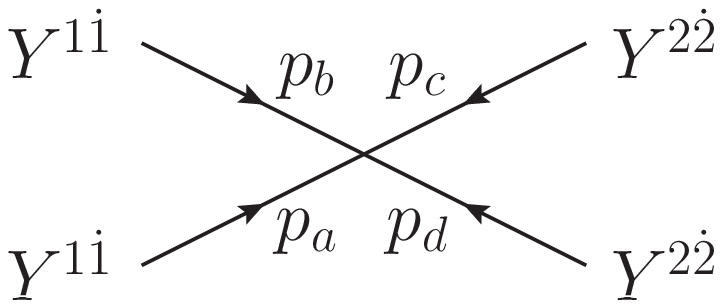}}}}&=\frac{i}{g}\big(c_1(p_a+p_b)^2+c_2(\omega_a+\omega_b)^2+2c_3\\
&\qquad+c_4[(p_a+p_c)^2+(p_a+p_d)^2]\nonumber\\
&\qquad+c_5[(\omega_a+\omega_c)^2+(\omega_a+\omega_d)^2]\big)\nonumber
\\[2pt]
\hbox{\raise-18pt\hbox{~\mbox{\includegraphics[height=17mm]{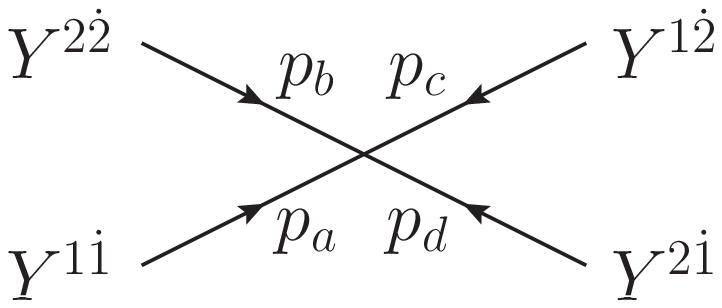}}}}&={}\frac{i}{g}\big(c_1(p_ap_b+p_cp_d)+c_2(\omega_a\omega_b+\omega_c\omega_d)-c_3\label{B-field pure Y}\\
&\qquad-c_4(p_a+p_b)^2-c_5(\omega_a+\omega_b)^2\nonumber\\
&\qquad+\beta_{12}(\omega_ap_b-\omega_bp_a)+\beta_{34}(\omega_cp_d-\omega_dp_c)\nonumber\\
&\qquad+\beta_{13}(\omega_ap_c-\omega_cp_a)\big)
\nonumber
\end{align}
for some choices of the constant coefficients $c_i$ and $\beta_{ij}$ which may be easily found by inspecting eqs.~\eqref{quarticeven}-\eqref{quarticodd}.

\begin{figure}[ht]
\begin{center}
\includegraphics[height=34mm]{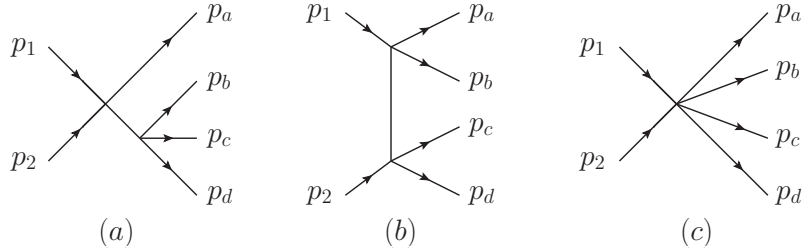}
\caption{Graph topologies contributing to the $2\rightarrow 4$ tree-level S-matrix element. 
One should include all possible assignments of outgoing momenta. \label{6pt_factorization}}
\end{center}
\end{figure}

We will consider explicitly the $2\rightarrow 4$ process with incoming fields  $Y^{1\dot{2}}$ and $Y^{2\dot{1}}$ with momenta $p_1$ and $p_2$, respectively;
for the outgoing fields we will take two $Y^{1\dot{1}}$s (with momenta $p_3$ and $p_4$) and two $Y^{2\dot{2}}$s (with momenta $p_5$ and $p_6$). The relevant 
Feynman graph topologies are shown in fig.~\ref{6pt_factorization}. 
The graph of type fig.~\ref{6pt_factorization}$(a)$ appears four times, where the outgoing leg with momentum $p_a$ can be assigned to any one of the 
outgoing fields. The graph of type fig.~\ref{6pt_factorization}$(b)$ appears in principle six times, with the outgoing legs with momenta $(p_a, p_b)$ being 
assigned to all possible pairs of momenta; due to our choice of flavor of outgoing fields however, two of such assignments ($(p_a, p_b)=(p_3, p_4)$
and $(p_a, p_b)=(p_5, p_6)$) vanish identically.

Straightforward algebra shows that upon using the identity
\begin{align}
\Delta^{-1}=&{}(\omega_a+\omega_b+\omega_c)^2-\alpha(p_a+p_b+p_c)^2-m^2,\\
(\omega_bp_c-\omega_c p_b)\Delta=&{}\frac{1}{4\alpha}\left(\frac{\omega_a-\omega_b}{p_a+p_b}-\frac{\omega_a-\omega_c}{p_a+p_c}\right)
\end{align}
and combining the eight contributions all propagators cancel out and
we find a local expression. For all choices of $c_i$ and $\beta_{ij}$ coefficients in \eqref{B-field pure Y} it can be put into a 
form reminiscent of the contribution of a six-point vertex:
\bea
i\tmatrix^{(0)}_{2\rightarrow 4}\Big|^{(a), (b)}_\text{parity-odd}
\!\!&=&\!\!
\frac{i}{4g^2}\left(\frac{c_1}{\alpha}-c_2\right)\bigg[2(6\beta_{34}+\beta_{13})\big(\omega_1p_2-\omega_2p_1\big)+(6\beta_{12}+\beta_{13})\big((\omega_3+\omega_4)(p_5+p_6)\cr
&&-(\omega_5+\omega_6)(p_3+p_4)\big)
+8\beta_{13}\big(\omega_1(p_3+p_4)-(\omega_3+\omega_4)p_1\big)\bigg]
\eea
It is not difficult to check that such a six-point vertex Feynman rule arises from the second term in the parity-odd six-field Wess-Zumino 
term in eq.~\eqref{WZ6pt}. 
We have also checked that the same is true for all parity-even and parity-odd six-point tree-level S-matrix elements.

\section{The one-loop S matrix \label{oneloop}}

A direct calculation of the one-loop S matrix is interesting for several reasons. On the one hand it would probe 
the integrability of the theory beyond classical level and it would determine to this order the dressing phase of the
S matrix (in the small momentum expansion). On the other it would explore the realization of 
symmetries at the quantum level and the extent to which the classical asymptotic states form a representation 
of the symmetry group assumed in the construction of the exact S matrix  \cite{Beisert:2008tw, beis, hhmphase}. Should the 
two realizations be different, an explicit expression of the S matrix in terms of classical asymptotic states would
allow us to determine the (nonlocal) redefinition that relates them to the true one-loop (and perhaps all-loop) 
states. We will denote henceforth this S matrix (and the corresponding $\tmatrix$ matrix) with the index "$b$".
%

In the following we will use unitarity-based methods  \cite{UnitarityMethod, BCFUnitarity} discussed in the context 
of two-dimensional integrable theories in \cite{Bianchi:2013nra, Engelund:2013fja, Bianchi:2014rfa} to find the one-loop and
the logarithmic terms of the two-loop S matrix. This construction will assume that the asymptotic states are the 
classical ones, with two-particle states realized as the tensor product of single-particle states.

An important ingredient in the construction of the S matrix through such methods are the tree-level 
S-matrix elements with fermionic external states, which are currently unknown from worldsheet methods. 
As  we shall see, to draw conclusions on the properties of asymptotic states only the general form of the 
tree-level S matrix and general properties of the tree-level coefficients (which may be justified by 
{\it e.g.} assuming integrability) are necessary. To find the actual expression of the loop-level S matrix 
we shall extract the tree-level fermionic S-matrix elements from the exact S matrix.

\subsection{Comments on unitarity vs. Feynman rules \label{finiteness}}

The construction of scattering matrices in two-dimensional integrable models  from unitarity cuts was discussed in detail
in  \cite{Bianchi:2013nra}, \cite{Engelund:2013fja}. While in \cite{Engelund:2013fja} only the terms with logarithmic momentum 
dependence were discussed,  ref.~\cite{Bianchi:2013nra} gave a prescription the calculation of the  complete one-loop S matrix;
it is interesting to discuss its relation to the Feynman diagram calculation in \cite{Roiban:2014cia} or the analogous calculation
in the $\eta$-deformed theory.

As discussed in \cite{Roiban:2014cia} in the context of undeformed AdS$_n\times$S$^n$ theories, the off-shell one-loop 
two-point function vanishes on shell. Moreover, the one-loop four-point function is also divergent and the on-shell 
divergence is proportional to the tree-level S matrix. The corresponding renormalization factors necessary to remove all
divergences are related to each other and can be simultaneously eliminated by a field redefinition. One may understand 
the relation between renormalization factors as a consequence of the (spontaneously broken) scale invariance of the theory.
Due to integrability, the unitarity-based calculation  \cite{Bianchi:2013nra}, \cite{Engelund:2013fja} is insensitive to the second 
type of divergence, which would correct the four-point interactions. Indeed, integrability in the form of the factorization of 
the six-particle amplitude implies that a one-particle cut of the one-loop four-point amplitude, which would identify the 
divergent tadpole integral, contains a further cut propagator and that it is in fact a two-particle cut and thus it predicts 
the absence of an infinite renormalization of the four-point vertex. 
This is consistent with the fact that the one-particle cut of the on-shell two-point function computed from the four-point 
S matrix vanishes as well. 
Thus, on the one hand, Feynman graph calculations exhibit divergences removable by field redefinitions while unitarity-based 
calculations are insensitive to any such divergences. 

Before embarking in the unitarity-based construction of the one-loop S matrix for the $\eta$-deformed theory, it is useful to 
check whether a similar consistent setup exists in this case as well. This is indeed the case. In the previous section we illustrated 
the fact that the six-point tree-level S matrix factorizes and thus the one-particle irreducible contributions to the one-loop
four-point S matrix are free of tadpole integrals. One can also check using the tree-level four-point S matrix \eqref{cTmatr} 
with coefficients given in Appendix~\ref{TreeSmatrix} that the one-particle cut of the on-shell one-loop two-point function 
vanishes as well.
Assuming that the worldsheet theory has indeed spontaneously-broken scale invariance (as it should to be a good worldsheet theory 
expanded around a nontrivial vacuum state) and by analogy with the undeformed case, we may therefore expect that unitarity-based 
calculation as described in \cite{Bianchi:2013nra}, \cite{Engelund:2013fja} will capture the complete one-loop S matrix.

\subsection{One-loop logarithmic terms  and the need for new asymptotic states}

To understand whether corrections to asymptotic states are necessary,  let us first construct the logarithmic terms 
of the one-loop S matrix under the standard assumption that the loop-level asymptotic states are the same 
as the tree-level ones and contrast the results with the consequences of integrability \eqref{SmatExpansion}. 

\begin{figure}[t]
\begin{center}
\includegraphics[height=32mm]{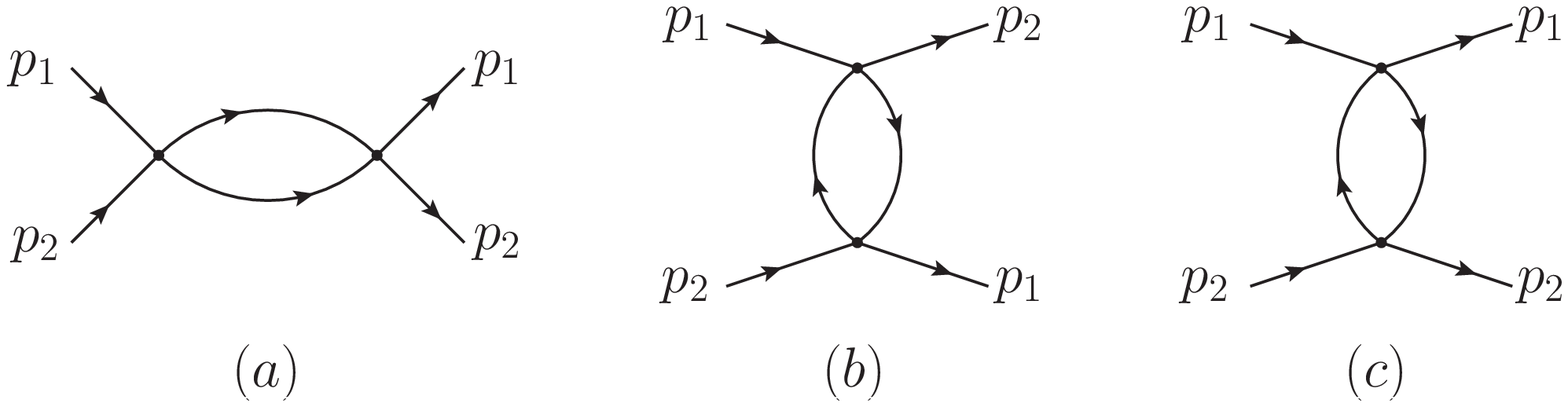}
\caption{The integrals appearing in the one-loop four-point amplitudes.  Tensor integrals can be reduced to them 
as well as to tadpole integrals, which are momentum-independent. \label{1loop_integrals}}
\end{center}
\end{figure}

\begin{figure}[t]
\begin{center}
\includegraphics[height=32mm]{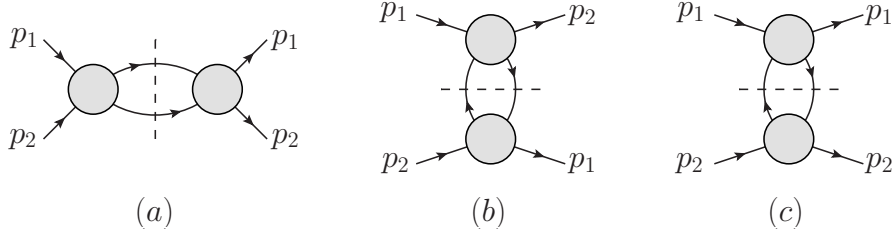}
\caption{Two-particle cuts of the one-loop four-point amplitudes \label{1loop_2p_cut}}
\end{center}
\end{figure}

To this end we use the unitarity-based method described in two-dimensional context 
in~\cite{Bianchi:2013nra, Engelund:2013fja, Bianchi:2014rfa}. 
The one-loop S matrix with tree-level asymptotic states (denoted by the lower index $b$) is given by
\bea
\label{1loopTmatrix}
i\tmatrix^{(1)}_b={}\tfrac{1}{2}C_sI_s+\tfrac{1}{2}C_uI_u+\tfrac{1}{2}C_tI_t     
={}\frac{i}{2\pi}\mathrm{ln}\left(\frac{p_{2-}}{p_{1-}}\right)\left(\frac{C_u}{J}-\frac{C_s}{J}\right)
-\frac{C_s}{2J}+\frac{i(1-\nu^2)^{3/2}}{8\pi}C_t \ .
\eea
where the integrals are shown in fig.~\ref{1loop_integrals}. We used their values for the propagators following 
from the action \eqref{action_quadratic}, and the $t$-channel integral was defined through Wick rotation to Euclidean 
space. The integral coefficients $C_s$, $C_u$ and $C_t$ are determined by unitarity cuts, with a suitable 
interpretation\footnote{To extract the coefficient of the $t$-channel one notices that on the one hand the formal 
cut of the $t$-channel integral is divergent due to the squared propagator and on the other the cut evaluated as a product
of tree-level amplitudes is also divergent due to the momentum conserving delta function. The prescription of \cite{Bianchi:2013nra}
is to identify the coefficients of these divergences in the limit in which the cut momentum equals one of the external momenta.}
of the singular $t$-channel cut:
\be
(C_s)_{AB}^{CD}  &=&(i)^2 J \sum_{E, F'}(i\tmatrix^{(0)})_{EF}^{CD} (i\tmatrix^{(0)})_{AB}^{EF}
\\[2pt]
(C_u)_{AB}^{CD}  &=& 
(i)^2J \sum_{E, F'}(-)^{([B]+[F])([D]+[F])}
(i\tmatrix^{(0)})_{EB}^{CF} (i\tmatrix^{(0)})_{AF}^{ED} 
\\[2pt]
(C_t)_{AB}^{CD}&=&{}(i)^2\sum_{E,F}(-)^{[E]([E]+[A])}\lim_{p_2\to p_1}\left(J(i\tmatrix^{(0)})_{AF}^{EC}\right)(i\tmatrix^{(0)})_{EB}^{FD}\nonumber \\
&=&{}(i)^2\sum_{E,F}(-)^{[F]([B]+[F])}(i\tmatrix^{(0)})_{AF}^{CE}\lim_{p_1\to p_2}\left(J(i\tmatrix^{(0)})^{DF}_{EB}\right) \ .
\label{generalC1loop}
\ee

Since the unitarity cuts fix completely the loop momentum, it is convenient to express the one-loop amplitude in terms 
of (the tree-level part of the) the coefficients $A,\dots,W$  parameterizing the S matrix, cf. eq.~\eqref{cTmatr}.
The Grassmann parity of states introduces relative signs between various contributions; to keep track of them it is convenient 
to introduce the parameter $\epsilon_{AB}$ with $A=1,\dots,4 \equiv \{a,\alpha\}$ defined as 
\be
\epsilon_{AB}=
\begin{cases}
\epsilon_{ab}  & A, B=1,2 \\
\epsilon_{\alpha\beta}  & A, B=3,4 \\
0      & A=1,2\; , \; \; B=3,4
\end{cases}
 \ .
\ee

The components of the difference between the $s$- and $u$-channel integral coefficients,
\be
\frac{C_u}{J}-\frac{C_s}{J} \ ,
\ee
expressed in terms of generic tree-level S-matrix coefficients  in eq.~\eqref{cTmatr} are collected in Appendix~\ref{CsMinusCu}. 
These expressions contain a variety of terms whose structure is different from that expected of the tensor part of the S matrix on the basis
of integrability and factorized symmetry. Assuming that the symmetry generators receive $1/g$ corrections, the only terms that may become 
consistent with symmetries are those proportional to the tree-level S matrix. Not all such terms survive however due to the identities
\begin{align}
A^{(0)}+D^{(0)}={}G^{(0)}+L^{(0)} \ ,  \qquad \quad B^{(0)}+E^{(0)}={}0 \ ,
\end{align}
which may be found using the expressions for the bosonic tree-level S-matrix elements found in \cite{Arutyunov:2013ega}. 
The terms that are not proportional to the tree-level S matrix must cancel; this requires that the following 
relation must hold:
\begin{align}
 (B^{(0)})^2+C^{(0)}F^{(0)}-H^{(0)}K^{(0)}-(W^{(0)}){}^2={}0 \ .
\end{align}
Showing that this holds requires knowledge of fermionic S-matrix elements. We extracted them from the exact S matrix 
of  \cite{Beisert:2008tw, beis, hhmphase}. Even 
though they have not yet been found through direct worldsheet calculations, the fact that the sigma model is classically 
integrable \cite{Delduc:2013qra} and has PSU$_q(2|2)^2$ quantum group symmetry \cite{Delduc:2014kha} suggests that 
they should be the correct ones.

Using these identities, eqs.~\eqref{ABcoef}-\eqref{Fcoef} can be compactly written as:
\begin{align}
\frac{C_s}{J}-\frac{C_u}{J}=&{}(H^{(0)}K^{(0)}+C^{(0)}F^{(0)})\mathbbm{1}+iW^{(0)}\left(\sum_{E=1}^4(\epsilon_{AE}-\epsilon^{CE})\right)(i\tmatrix^{(0)}).
\label{forskel cuts 1-loop}
\end{align}
Thus, it follows that the logarithmic terms of the one-loop S matrix with tree-level asymptotic states are given by
%
%
\begin{align}
i\tmatrix_b^{(1)}\Big|_{\ln\;\text{terms}}=&\frac{1}{2\pi}W^{(0)}\left(\sum_{E=1}^4(\epsilon_{AE}-\epsilon^{CE})\right)
\mathrm{ln}\left(\frac{p_{2-}}{p_{1-}}\right)(i\tmatrix^{(0)}) \nonumber\\
&{}-\frac{i}{2\pi}(H^{(0)}K^{(0)}+C^{(0)}F^{(0)})\mathrm{ln}\left(\frac{p_{2-}}{p_{1-}}\right)\mathbbm{1} \ .
\label{Resultat_1-loop_logs}
\end{align}
We note that the first line of this expression is inconsistent with the expansion \eqref{SmatExpansion}  of the S matrix suggested by quantum integrability 
and the expected PSU$_q(2|2)^2$ symmetry.  Indeed, eq.~\eqref{SmatExpansion} implies that at one-loop level the only logarithmic 
momentum dependence appears in the S matrix phase -- and thus the only logarithms multiply the unit operator -- while the tensor part 
is free of logarithms.

\subsection{One-loop symmetries and new asymptotic two-particle states \label{sec:1loopStrue}}

The fact that the offending term in eq.~\eqref{Resultat_1-loop_logs} is proportional to the tree-level S matrix suggests 
that it should be possible to eliminate it by a redefinition of the asymptotic states. At tree level 
these states are tensor product of single-particle states however this does not need to be the case at loop level. 
We will consider two redefinitions:  
$(a)$ one makes the spin and dimension of the single-particle states complex while preserving the tensor-product structure 
of the two-particle state and the other
$(b)$ does not act independently on the single-particle states but breaks the tensor product of the two-particle states. 
While distinct, the two redefinitions have the same effect on the S matrix and put it in a form consistent with the consequences 
of integrability and expected symmetries. 

To identify the desired transformation we notice that for all choices of external states the following identity holds:
\begin{align}
\sum_{E=1}^4(\epsilon_{AE}-\epsilon^{CE})=-\sum_{E=1}^4(\epsilon_{BE}-\epsilon^{DE}) \ .
\label{epidentity}
\end{align}
For diagonal elements, $A=C$ and $B=D$, 
both the left-hand and the right-hand side are trivially zero, while they are non-vanishing for off-diagonal S-matrix elements.
Using this identity,  the two possible redefinitions are:
\bea
\label{1Pstate_redef}
(a)&~~~& 
\begin{aligned} 
|A,p\rangle\mapsto p_-^{+\frac{W^{(0)}}{2\pi g}\sum_{E=1}^4\epsilon_{AE}}|A,p\rangle \ ,\cr
 \langle C,p|\mapsto p_-^{-\frac{W^{(0)}}{2\pi g}\sum_{E=1}^4\epsilon^{CE}}\langle C,p|  \ ;
 \end{aligned}
\\ 
\label{2Pstate_redef}
(b) &~~~&
\begin{aligned} 
|A, p_1\rangle \otimes |B, p_2\rangle \mapsto   e^{+\frac{W^{(0)}}{4\pi g}\mathrm{ln}\left(\frac{p_{1-}}{p_{2-}}\right)\sum_{E=1}^4(\epsilon_{AE}-\epsilon_{BE})}  
|A, p_1\rangle \otimes |B, p_2\rangle \ ,
\cr
\langle C, p_1| \otimes \langle D, p_2|   \mapsto   e^{-\frac{W^{(0)}}{4\pi g}\mathrm{ln}\left(\frac{p_{1-}}{p_{2-}}\right)\sum_{E=1}^4(\epsilon^{CE}-\epsilon^{DE})}  
\langle C, p_1| \otimes \langle D, p_2| \ .
\end{aligned}
\eea
Since the tree-level coefficient $W^{(0)}$ is purely imaginary, $W^{(0)}{}^* = -W^{(0)}$, 
cf. eq.~\eqref{Wtree}, both redefinitions preserve the unitarity properties of the original S matrix as 
the in and out states remain hermitian conjugates of each other.
We also notice that the redefinition $(b)$ is not sensitive to the order of the 
states in the original tensor product. Of course, at one loop only the first term in the expansion of the 
exponential factors is relevant; we however keep the full exponential form to exhibit manifest unitarity of the
state transformation. 

In terms of the new  asymptotic states and upon using eq.~\eqref{epidentity} the one-loop S matrix becomes
\begin{align}
\label{1loopSfull}
i\tmatrix^{(1)} 
=&{}\frac{i}{2\pi}\mathrm{ln}\left(\frac{p_{2-}}{p_{1-}}\right)\left(\frac{C_u}{J}-\frac{C_s}{J}+iW^{(0)} \sum_{E=1}^4(\epsilon_{AE}-\epsilon^{CE})(i\tmatrix^{(0)})\,\right)
-\frac{C_s}{2J}+\frac{i(1-\nu^2)^{3/2}}{8\pi}C_t \ ;
\end{align}
by construction the logarithmic terms proportional to $(i\tmatrix^{(0)})$ cancel in the parenthesis and we are left with an expression consistent with 
integrability and expected symmetries. In the limit of vanishing deformation parameter, $\nu\rightarrow 0$, the bare and redefined states become 
identical, as required by the fact that no state redefinition is necessary in the undeformed theory.

Following \cite{Bianchi:2013nra}, the $t$-channel integral coefficient $C_t$ can be found by removing the vanishing Jacobian factor 
from the tree-level S matrix, eq.~\eqref{generalC1loop}, and is given by:
\begin{align}
\label{Ct1loop}
C_t
=&{}\frac{4}{1-\nu^2}\frac{(\omega_1^2-1)(\omega_2^2-1)}{\omega_2 p_1-p_2\omega_1}\mathbbm{1} \ .
\end{align}
In the limit of zero deformation this coefficient gives rise to the rational part of the one-loop dressing phase whereas $C_s$ gives the one-loop terms in the expansion of the coefficients $A,\dots,K$ in the definition 
\eqref{cTmatr} of the S matrix.\footnote{In theories with cubic 
interaction terms there may exist nontrivial corrections to the two-point function of fields which change its residue at the 
physical pole. This leads to further terms in the one-loop S matrix, see \cite{Bianchi:2014rfa}. The $\eta$-deformed AdS$_5\times$S$^5$ 
Lagrangian has only quartic (and higher-point) vertices and thus such corrections appear only at two loops.}
For non-vanishing $\eta$-parameter we have checked that this continues to be the case by comparing the entries of $C_s$ with the perturbative expansion
of the exact S-matrix coefficients~ \cite{Beisert:2008tw, beis, hhmphase}.
%
We collect
the expressions of the one-loop S-matrix coefficients in Appendix~\ref{oneloopSmatrixcoefs}.

\section{The two-loop S matrix and consistency of the asymptotic states \label{twoloop} }

In \cite{Engelund:2013fja} the double-logarithms of the two-loop S matrix were computed from double two-particle
cuts and expressed in terms of two-loop scalar integrals.  Additional single-logarithms were then found from 
single two-particle cuts, making use of the rational part of the one-loop S matrix determined by 
symmetries \footnote{The part proportional to the identity operator cancelled out.}. The result was, however, expressed 
only in terms of one-loop integrals. Here we identify a particular set of two-loop scalar and tensor integrals which allows us to write 
a uniform two-loop integral representation of all two-loop logarithmic terms. 

\subsection{A set of tensor integrals}

\begin{figure}[ht]
\begin{center}
\includegraphics[height=68mm]{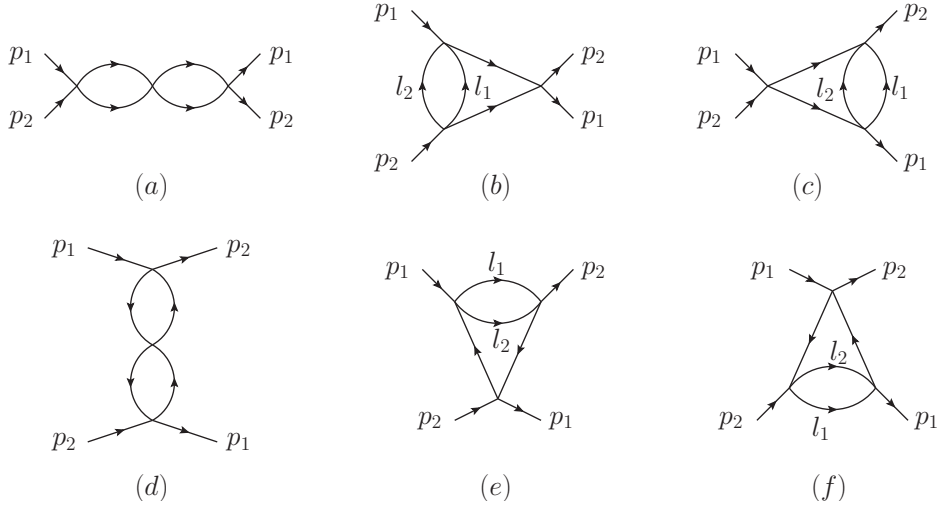}
\caption{The integrals appearing in the two-loop four-point amplitudes.  \label{2loop_integrals}}
\end{center}
\end{figure}

The topologies of the contributing integrals are the same as in \cite{Engelund:2013fja} and are shown in 
fig.~\ref{2loop_integrals}. Let us parametrize the integrals in figures \ref{2loop_integrals}$(b)$, $(c)$, $(e)$ 
and $(f)$ as shown in the figure. We first consider the cut in the $s=(p_1+p_2)^2$ channel, which receives contributions
from graphs with topologies $(a)$, $(b)$ and $(c)$.
If one interprets them as scalar integrals, then
the two-particle cut condition for the graph $(a)$ has two solutions
both of which are proportional to the $s$-channel one-loop integral. The two particle cut conditions of graphs $(b)$ 
and $(c)$ also have two solutions; however, one of them is proportional to the $t$-channel one-loop integral while 
the other is proportional to the $u$-channel one; the corresponding solutions may be parameterized as
\be
l_1+l_2 = 0
\quad
\text{and}
\quad
l_1+l_2 = p_2 - p_1 \ .
\label{cc}
\ee 
Instead of using them however, we shall define  integrals whose single two-particle cuts receive contributions 
from a single one-loop integral. This can be easily 
done by making use of the solution \eqref{cc} to the cut condition and inserting appropriate momentum-dependent 
numerator factors. Denoting by $D_{a,b, c, d,e, f}$ the denominator of the products of scalar propagators corresponding to the graphs in
fig.~\ref{2loop_integrals}, they are
\bea
I_a=\int \frac{d^2l_1d^2l_2}{(2\pi)^4} \frac{1}{D_a} 
&& 
I_d=\int \frac{d^2l_1d^2l_2}{(2\pi)^4} \frac{1}{D_d}
\\
I_b=\int \frac{d^2l_1d^2l_2}{(2\pi)^4} \frac{n_b}{D_b} 
&& 
I_c=\int \frac{d^2l_1d^2l_2}{(2\pi)^4} \frac{n_c}{D_c}
\quad 
I_e=\int \frac{d^2l_1d^2l_2}{(2\pi)^4} \frac{n_e}{D_e}
\quad
 I_f=\int \frac{d^2l_1d^2l_2}{(2\pi)^4} \frac{n_f}{D_f}
 \\
 I_g=\int \frac{d^2l_1d^2l_2}{(2\pi)^4} \frac{n_g}{D_b} 
&& 
I_h=\int \frac{d^2l_1d^2l_2}{(2\pi)^4} \frac{n_h}{D_c}
\quad 
I_k=\int \frac{d^2l_1d^2l_2}{(2\pi)^4} \frac{n_k}{D_e}
\quad
 I_l=\int \frac{d^2l_1d^2l_2}{(2\pi)^4} \frac{n_l}{D_f} \ ,
\eea
where
\begin{align}
n_b=n_c&=\frac{l_1+l_2}{p_2-p_1} \ ,&n_e=n_f&=\frac{l_1+l_2}{p_2+p_1} \ ,\\
n_g=n_h&=1-\frac{l_1+l_2}{p_2-p_1} \ ,&n_k=n_l&=1-\frac{l_1+l_2}{p_2+p_1} \ .
\end{align}
The $s$-channel two-particle cut of the integrals $I_b$ and $I_c$ receives contributions only from $u$-channel 
one-loop sub-integrals, the $u$-channel two-particle cut of the integrals $I_e$ and $I_f$ receives contributions 
only from $s$-channel one-loop sub-integrals while the two-particle cuts of the remaining integrals  receive 
contributions only from the $t$-channel one-loop sub-integrals.

In terms of the ten scalar and tensor integrals $I_a,\dots, I_l$, the ansatz for the logarithmic terms of the two-loop S matrix is
\begin{align}
\label{Ansatz 2-loop}
i\tmatrix^{(2)}_b=&{}\frac{1}{4}C_aI_a+\frac{1}{4}C_dI_d
+\frac{1}{2}C_bI_b+\frac{1}{2}C_cI_c
+\frac{1}{2}C_eI_e+\frac{1}{2}C_fI_f
\cr
&\qquad\qquad\qquad\;\;\;
+\frac{1}{2}C_gI_g+\frac{1}{2}C_hI_h
+\frac{1}{2}C_kI_k+\frac{1}{2}C_lI_l
+\text{extra rational terms} \ ;
\end{align}
using the explicit expressions for the integrals listed in Appendix~\ref{integrals} it is not difficult to see that 
$i\tmatrix_b^{(2)}$ may be written as
\begin{align}
i\tmatrix^{(2)}_b=&{}\frac{1}{8\pi^2J^2}\mathrm{ln}^2\left(\frac{p_{2-}}{p_{1-}}\right)\left(-2C_a+C_b+C_c-2C_d+C_e+C_f\right)
\label{2-loop S matrix without t}\\
&+\frac{i}{2\pi}\mathrm{ln}\left(\frac{p_{2-}}{p_{1-}}\right)\left[\frac{1}{2J^2}(2C_a-C_b-C_c)-\frac{i(1-\nu^2)^{3/2}}{16\pi J}(C_g+C_h-C_k-C_l)\right]\nonumber\\
&+\frac{1}{4J^2}C_a+\frac{(1-\nu^2)\omega_1\omega_2-p_1p_2-1}{16J^2}(C_b+C_c-C_e-C_f)\nonumber\\
&-\frac{1}{4J}\left(\frac{i(1-\nu^2)^{3/2}}{4\pi}\right)(C_g+C_h)\nonumber\\
&+\text{extra rational terms}\nonumber \ .
\end{align}
The ten coefficients $C_a,\dots,C_l$ are determined by {\em single} two-particle cuts in terms of 
tree-level amplitudes and one-loop integral coefficients $C_s$, $C_u$ and $C_t$. Each solution to the cut condition 
determines exactly one coefficient. The first six coefficients have the same expression as the coefficients with the same 
name in ref.~\cite{Engelund:2013fja},
\bea
(C_a){}_{AB}^{CD}&=&{}(i)^2J\sum_{G,H}(i\tmatrix^{(0)})_{GH}^{CD}(C_s)_{AB}^{GH}
={}(i)^2J_s\sum_{G,H}(C_s)_{GH}^{CD}(i\tmatrix^{(0)})_{AB}^{GH}\nonumber\\[2pt]
(C_b){}_{AB}^{CD}&=&{}(i)^2J \sum_{G,H}(i\tmatrix^{(0)})_{GH}^{CD}
(C_u)_{AB}^{GH}\nonumber\\[2pt]
(C_c){}_{AB}^{CD}&=&{}(i)^2J \sum_{G,H}(C_u){}_{GH}^{CD} (i\tmatrix^{(0)})_{AB}^{GH}
\nonumber\\[2pt]
(C_d){}_{AB}^{CD}&=&{}(i)^2J \sum_{G,H}   
(-)^{([B]+[H])([D]+[H])} (i\tmatrix^{(0)})_{GB}^{CH} (C_u){}_{AH}^{GD}
\nonumber\\[2pt]
&=&{}(i)^2J \sum_{G,H}   
(-)^{([B]+[H])([D]+[H])} (C_u)_{GB}^{CH} (i\tmatrix^{(0)}){}_{AH}^{GD}
\nonumber\\[2pt]
(C_e){}_{AB}^{CD}&=&{}(i)^2J\sum_{G,H} (-)^{([B]+[H])([D]+[H])} (i\tmatrix^{(0)}){}_{GB}^{CH} 
(C_s){}_{A H}^{G D}
\nonumber\\[2pt]
(C_f){}_{AB}^{CD}&=&{}(i)^2J\sum_{G,H}   (-)^{([B]+[H])([D]+[H])}(C_s){}_{GB}^{CH}
(i\tmatrix^{(0)}){}_{AH}^{GD} \ ,
\eea
while  $C_g$, $C_h$, $C_k$ and $C_l$ are given by
\begin{align}
C_g=&{}(i)^2J\sum_{G,H}(i\tmatrix^{(0)})_{GH}^{CD}(C_t)_{AB}^{GH} \ ,\quad
C_h={}(i)^2J\sum_{G,H}(C_t)_{GH}^{CD}(i\tmatrix^{(0)})_{AB}^{GH} \ ,\\
C_k=&{}(i)^2J \sum_{G,H}   
(-)^{([B]+[H])([D]+[H])} (i\tmatrix^{(0)})_{GB}^{CH} (C_t){}_{AH}^{GD}\nonumber\\
C_l=&{}(i)^2J \sum_{G,H}  \,
(-)^{([B]+[H])([D]+[H])} (C_t)_{GB}^{CH} (i\tmatrix^{(0)}){}_{AH}^{GD}
\nonumber \ .
\end{align}

Since the coefficient of the one-loop logarithms depends only on the differences $(C_s-C_u)$ (cf. eq.~\eqref{1loopTmatrix}), 
the coefficient of the two-loop double-logarithm should have a similar property. As discussed in the previous section, this difference has two parts;
one proportional to the identity operator and one proportional to the tree-level S matrix. It is not difficult to check 
that the part proportional to the identity operator cancels 
out in the two-loop S matrix; the remaining bilinear in tree-level S-matrix elements can again be organized in terms of the difference 
$(C_s-C_u)$.  Upon using the identity 
\begin{align}
\left(\sum_{E=1}^4(\epsilon_{AE}-\epsilon^{CE})\right)\mathbbm{1}=&{}0 
\label{ep_identity}
\end{align}
the coefficient of the double-logarithm becomes proportional to the tree-level S matrix, and may be suggestively organized as
\begin{align}
(i\tmatrix^{(2)}_b)_{AB}^{CD}
=&{}\frac{1}{2!}\left(\frac{W^{(0)}}{2\pi}\right)^2\left(\sum_{E=1}^4(\epsilon_{AE}-\epsilon^{CE})\right)^2\mathrm{ln}^2\left(\frac{p_{2-}}{p_{1-}}\right)(i\tmatrix^{(0)})
+\text{single log}+\text{rational} \ .
\end{align}

To find the coefficient of the simple logarithms we first recall \cite{Engelund:2013fja} that, on general grounds related to the consistency of 
single and double two-particle cuts of two-loop S-matrix elements,  the contribution of terms proportional to the identity operator in the 
one-loop S matrix vanishes. 
Using eq.~\eqref{ep_identity} we find that, up to rational terms, the two-loop S matrix is given by
\begin{align}
i\tmatrix^{(2)}_b=&-\frac{1}{2!}\left(\frac{W^{(0)}}{2\pi}\right)^2\left(\sum_{E=1}^4(\epsilon_{AE}-\epsilon^{CE})\right)^2\mathrm{ln}^2\left(\frac{p_{2-}}{p_{1-}}\right)
(i\tmatrix^{(0)})
\nonumber\\
&{}+\frac{W^{(0)}}{2\pi}\left(\sum_{E=1}^4(\epsilon_{AE}-\epsilon^{CE})\right)\mathrm{ln}\left(\frac{p_{2-}}{p_{1-}}\right)(i\tmatrix^{(1)}_b)\\
&{}-\frac{i}{2\pi}(H^{(0)}K^{(0)}+C^{(0)}F^{(0)})(i\tmatrix^{(0)})\mathrm{ln}\left(\frac{p_{2-}}{p_{1-}}\right)+\text{rational} \ .
\nonumber
\end{align}

As in the case of the one-loop S matrix, this expression is not immediately consistent with the implications of symmetries and integrability \eqref{SmatExpansion}.
Using however the one-loop corrected asymptotic states, all offending terms cancel out and we find
\begin{align}
i\tmatrix^{(2)}=&{}-\frac{i}{2\pi}(H^{(0)}K^{(0)}+C^{(0)}F^{(0)})\mathrm{ln}\left(\frac{p_{2-}}{p_{1-}}\right)\, (i\tmatrix^{(0)})+\text{rational}.
\end{align}
This is indeed the expected structure of the two-loop S matrix. Thus, the exponentiation of the one-loop redefinition of the asymptotic 
states \eqref{1Pstate_redef}, \eqref{2Pstate_redef} does not receive further two-loop 
corrections. It is natural to conjecture that the same holds at higher loops as well; it would, of course, be interesting to 
verify whether this is indeed the case.

\subsection{Comments on rational terms\label{wild}}

\begin{figure}[t]
\begin{center}
\includegraphics[height=37mm]{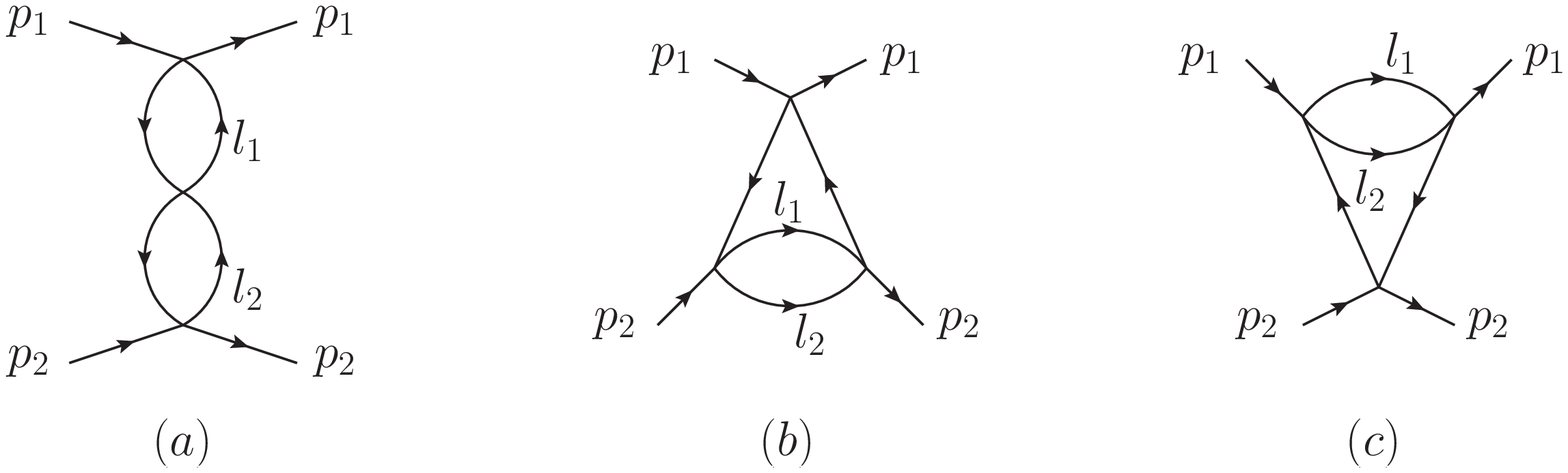}
\caption{Two-loop integrals which should provide the missing two-loop rational terms in four-point amplitudes.  \label{2loop_rational}}
\end{center}
\end{figure}

The combination of two-loop integrals \eqref{Ansatz 2-loop} giving the correct single and double-logarithms also contains 
some rational terms originating from the rational terms in the expressions of the two-loop integrals \eqref{2loopints}. By construction 
however, these terms do not account for all the possible two-particle cuts of the two-loop S matrix, in particular the cuts in which 
there is no net momentum flow across it,  which are analogous to the one-loop $t$-channel cut;
the potentially missing relevant integral topologies are shown in fig.~\ref{2loop_rational}. As in that case, one can convince 
oneself that all integrals based on these graphs 
are momentum independent (and thus their cuts are to be understood in a formal 
sense) and consequently they can contribute {\em only} rational terms to the two-loop S matrix. 
A further source of rational terms are the quantum corrections to the off-shell two-point function and additional integrals that have 
only $t$-channel two-particle cuts. 

\begin{figure}[t]
\begin{center}
\includegraphics[height=23mm]{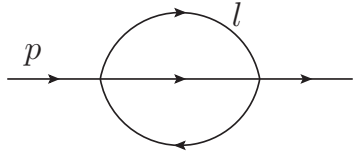}
\caption{Sunset diagrams are responsible for corrections to the two-point function and affect the two-loop S matrix\label{QC prop}}
\end{center}
\end{figure}

%
The  first corrections to the two-point function of fields arise from Feynman graphs of 
topology \footnote{In principle there are also graphs containing tadpoles, but they 
should cancel out and the final contribution arises effectively only from the topology 
in fig.~\ref{QC prop}.} \footnote{The fact that the first correction appears at two-loop level 
follows from the absence of cubic vertices in the gauge-fixed Lagrangian. In theories 
where such cubic vertices are present the first correction appears already at one loop, see {\it e.g.} \cite{Bianchi:2014rfa}.} 
shown in fig.~\ref{QC prop} and change the residue of the 
propagator at the physical pole; this must be accounted for in the definition of the 
S matrix.  Since the first correction to the dispersion relation arises at two-loop order, 
the additional terms in the two-loop S matrix are necessarily proportional to the 
tree-level S matrix. 
%

In the following we will not determine all rational terms; rather, we will point out specific features which appear to suggest how they can be found through generalized unitarity.  We begin by pointing out an interesting property of the calculation of the two-loop S 
matrix \cite{Klose:2007rz} in the  near-flat space limit \cite{Maldacena:2006rv} of AdS$_5\times$S$^5$. In this limit all integrals with 
the topology in fig.~\ref{2loop_rational}(a) have vanishing coefficients, while the integrals with the topology in figs.~\ref{2loop_rational}(b) 
and \ref{2loop_rational}(c) exactly cancel the quantum corrections to the external states. 
We will attempt to show that a similar pattern may be realized in general; we will also see that for this to happen it is necessary that 
the integral representation of the two-loop S matrix contains integrals that do not have two-particle cuts.

Since, as mentioned earlier, the two-loop corrections to the two-point function contribute to the two-loop S matrix terms proportional 
to the tree-level S matrix, to check the fate of these terms we shall focus on the integral coefficients corresponding to the 
topologies shown in fig.~\ref{2loop_rational} which are also  proportional to the  tree-level S matrix. 
There are six combinations of tree-level S-matrix elements and one-loop integral coefficients 
which can appear in two-particle cuts (and thus determine these integrals' coefficients) and have this property:
\begin{align}
(X_a)_{AB}^{CD}=&{}(i)^2\sum_{E,F}(-)^{[E]([E]+[A])}\lim_{p_2\to p_1}\left(J(C_u)_{AF}^{EC}\right)(i\tmatrix^{(0)}){}_{EB}^{FD}\nonumber\\
(X_b)_{AB}^{CD}=&{}(i)^2\sum_{E,F} (-)^{[F]([B]+[F])}(i\tmatrix^{(0)})_{AF}^{CE}\lim_{p_1\to p_2}\left(J(C_t)^{DF}_{EB}\right)\label{rational cuts 2-loop}\\
(X_c)_{AB}^{CD}=&{}(i)^2\sum_{E,F}(-)^{[E]([E]+[A])}\lim_{p_2\to p_1}\left(J(C_t)_{AF}^{EC}\right)(i\tmatrix^{(0)}){}_{EB}^{FD}\nonumber\\
(X_d)_{AB}^{CD}=&{}(i)^2\sum_{E,F} (-)^{[F]([B]+[F])}(i\tmatrix^{(0)})_{AF}^{CE}\lim_{p_1\to p_2}\left(J(C_u)^{DF}_{EB}\right)\nonumber\\
(X_e)_{AB}^{CD}=&{}(i)^2\sum_{E,F} (-)^{[F]([B]+[F])}(i\tmatrix^{(0)})_{AF}^{CE}\lim_{p_1\to p_2}\left(J(C_s)^{DF}_{EB}\right)\nonumber\\
(X_f)_{AB}^{CD}=&{}(i)^2\sum_{E,F}(-)^{[E]([E]+[A])}\lim_{p_2\to p_1}\left(J(C_s)_{AF}^{EC}\right)(i\tmatrix^{(0)}){}_{EB}^{FD}\nonumber \ .
\end{align}
It is not difficult to identify these combinations as contributions to two-particle cuts from a single solution to the cut condition.
The other solutions contribute terms proportional to the identity matrix and, while important for the complete S matrix 
(in particular for the determination of the complete two-loop dressing phase) will ignored in the following. 
The numerator factors which can be used to dress the graphs in fig.~\ref{2loop_rational} and select the desired solution to
the two-particle cut condition such that the resulting integrals have $X_{a,\dots,f}$ as coefficients are:
\begin{align}
n_a^R=&{}\left(1-\frac{l_{2-}}{p_{2-}}\right)\left(1+\frac{l_{2-}}{p_{2-}}\right)\nonumber\\
n_b^R=&{}\left(1-\frac{l_{1-}+l_{2-}}{p_{1-}+p_{2-}}\right)\left(1+\frac{l_{1-}+l_{2-}}{p_{1-}-p_{2-}}\right)\left(1-\frac{l_{1-}+l_{2-}}{2p_{2-}}\right)\nonumber\\
n_c^R=&{}\left(1-\frac{l_{1-}+l_{2-}}{p_{1-}+p_{2-}}\right)\left(1-\frac{l_{1-}+l_{2-}}{p_{1-}-p_{2-}}\right)\left(1-\frac{l_{1-}+l_{2-}}{2p_{1-}}\right)\label{2-loop n rational}\\
n_d^R=&{}\left(1-\frac{l_{1-}}{p_{1-}}\right)\left(1+\frac{l_{1-}}{p_{1-}}\right)\nonumber\\
n_e^R=&{}\left(1-\frac{l_{1-}+l_{2-}}{p_{1-}+p_{2-}}\right)\left(1+\frac{l_{1-}+l_{2-}}{p_{1-}-p_{2-}}\right)\frac{l_{1-}+l_{2-}}{2p_{2-}}\nonumber\\
n_f^R=&{}\left(1-\frac{l_{1-}+l_{2-}}{p_{1-}+p_{2-}}\right)\left(1-\frac{l_{1-}+l_{2-}}{p_{1-}-p_{2-}}\right)\frac{l_{1-}+l_{2-}}{2p_{1-}} \ .\nonumber
\end{align}
Then, denoting as before by $D_{a,b,c}^{R}$ the denominators of scalar propagators associated to the graphs in fig.~\ref{2loop_rational}, the 
integrals whose coefficients are given by eqs.~\eqref{rational cuts 2-loop} are\footnote{The reader may notice that the systematic for generalizing the 
one-loop cuts to two-loop cuts seems to have been flipped around for the $t$-channel. This is indeed the case as a careful analysis along the lines 
of \cite{Bianchi:2013nra} will show.} :
\begin{align}
R_a&=\int \frac{d^2l_1d^2l_2}{(2\pi)^4} \frac{n_a^R}{D_a^R} 
&
R_b&=\int \frac{d^2l_1d^2l_2}{(2\pi)^4} \frac{n_b^R}{D_b^R} 
&
R_c&=\int \frac{d^2l_1d^2l_2}{(2\pi)^4} \frac{n_c^R}{D_c^R} 
\\
R_d&=\int \frac{d^2l_1d^2l_2}{(2\pi)^4} \frac{n_d^R}{D_a^R} 
&
R_e&=\int \frac{d^2l_1d^2l_2}{(2\pi)^4} \frac{n_e^R}{D_b^R}
&
R_f&=\int \frac{d^2l_1d^2l_2}{(2\pi)^4} \frac{n_f^R}{D_c^R} \ .\nonumber
\end{align}
We choose to use light-like directions in the numerator factors only for convenience, the resulting integrals having been already 
computed in \cite{Klose:2007rz}; a different choice would lead to different values for the integrals. 
%
It is not surprising that different numerator factors are possible: indeed, by expressing all loop momenta in terms of external 
momenta in two dimensions, cuts cannot determine 
unambiguously the tensor structure of an integral. Interestingly, the coefficients $X_{a,\dots,f}$ are such that when 
the component of the loop momenta used to construct the numerator is changed, the extra terms in the two-loop S matrix are 
proportional to the scalar sunset integral, see fig.~\ref{QC prop}.

To compute the corrections to external states the off-shell two-point function is necessary, because the residue 
of the corrected propagator contains the derivative of the two-point function with respect to the worldsheet energy. 
We shall {\em assume} that this derivative is entirely given by the derivative of  the integral. With this assumption 
we shall construct the two-point function by sewing two legs of the one-loop S matrix. There are four possible index contractions:
\bea
\begin{aligned}
&(X_g)_{A}^{C}={}\sum_E(-)^{[E]([C]+[E])}\lim_{p_2\to p_1}(J(C_s)_{AE}^{EC}) \ ,
&(X_h)_{A}^{C}={}\sum_E(-)^{[E]([C]+[E])}\lim_{p_2\to p_1}(J(C_u)_{AE}^{EC}) \ ,\\
&(X_k)_{B}^{D}={}\sum_E(-)^{[F]([B]+[F])}\lim_{p_1\to p_2}(J(C_s)_{FB}^{DF}) \ ,
&(X_l)_{B}^{D}={}\sum_E(-)^{[F]([B]+[F])}\lim_{p_1\to p_2}(J(C_u)_{FB}^{DF}) \ ,
\end{aligned}
\eea
\noindent 
which are the direct two-loop generalizations of the one-loop contractions 
\begin{align}
\sum_E(-)^{[E]([C]+[E])}\lim_{p_2\to p_1}(J(i\tmatrix^{(0)})_{AE}^{EC})
\quad,\quad
\sum_E(-)^{[F]([B]+[F])}\lim_{p_1\to p_2}(J(i\tmatrix^{(0)})_{FB}^{DF}),
\end{align}
\noindent 
which indicate the absence of external line tadpoles in the one-loop S matrix. 

The contractions $X_{g,\dots,l}$ are coefficients of  integrals with the topology given in 
fig.~\eqref{QC prop} whose one-particle cuts localize on a single solution of the cut condition.
Denoting by  $D_g$ and $D_h$ the denominators of the product of scalar propagators corresponding to the graph 
fig.~\eqref{QC prop} with external momentum $p_1$ and $p_2$, 
respectively, the numerator factors $n_{g,h,k,l}^R$ that lead to the desired localization are
\be
n_g^R={}\frac{1}{2}\left(1-\frac{l}{p_1}\right)
~~,\quad
n_h^R={}\frac{1}{2}\left(1+\frac{l}{p_1}\right)
~~,\quad
n_k^R={}\frac{1}{2}\left(1-\frac{l}{p_2}\right)
~~,\quad
n_l^R={}\frac{1}{2}\left(1+\frac{l}{p_2}\right)  \ ,
\ee
\noindent 
and lead to the integrals
\be
R_{g, h}=\int \frac{d^2l_1d^2l_2}{(2\pi)^4} \frac{n_{g, h}^R}{D_g^R}
\quad,\quad
R_{k,l}=\int \frac{d^2l_1d^2l_2}{(2\pi)^4} \frac{n_{k,l}^R}{D_k^R} \ .
\ee
%
It turns out that the result also depends on the scalar sunset integral; we shall denote this integral by
\begin{align}
R_0=&{}\int\frac{d^2l_1d^2l_2}{(2\pi)^4} \frac{1}{D_g^R} \ .
\end{align}

%

Putting all this together the several contributions, we get that the additional rational terms proportional to the tree-level S matrix 
not already present in \eqref{2-loop S matrix without t} can be written as
\begin{align}
\delta(i\tmatrix_b)^{\propto \tmatrix^{(0)}}=&{} \frac{1}{4}X_aR_a+\frac{1}{2}X_bR_b+\frac{1}{2}X_cR_c+\frac{1}{4}X_dR_d+\frac{1}{2}X_eR_e+\frac{1}{2}X_fR_f\label{extra1}\\
&+\left(\frac{1}{6}X_gR_g'+\frac{1}{6}X_hR_h'+\frac{1}{6}X_kR_k'+\frac{1}{6}X_lR_l'\right) (i\tmatrix^{(0)})
\end{align}
\noindent 
where the primes indicate derivative with respect to the time-like component of the external momentum.

\begin{figure}[ht]
\begin{center}
\includegraphics[height=22mm]{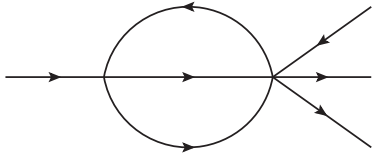}
\caption{The graph with no two-particle cuts that is expected to appear in the two-loop S matrix to restore the factorization of the 
three-particle cut required by tree-level integrablity. \label{6pt-vertex}}
\end{center}
\end{figure}

It is straightforward to calculate all the integral coefficients. We find that $X_a=0=X_d$ impliying that, similarly with 
the near-flat space calculation \cite{Klose:2007rz}, the two-point function depends only on integrals of wineglass topology. 
As some of the non-zero coefficients are equal the results can be written in terms of integrals simpler than the integrals in the 
basis, the useful combinations are collected in appendix \ref{integrals}. The result can be written as:
\begin{align}
\delta(i\tmatrix_b)^{\propto \tmatrix^{(0)}}=&{}(i\tmatrix^{(0)}) \; f(p_1,p_2) \, R_0 \ .
\end{align}
We have not explicitly written out the function $f(p_1,p_2)$ because it can be changed by changing the momentum 
components used in
%
%
the numerators in \eqref{2-loop n rational}. We notice that the remaining term is proportional to the integral $R_0$ which is 
associated with topologies of  the type shown in fig.~\ref{6pt-vertex}. 
Even though $R_0$ is constant, this is not a vacuous statement: while we have not determined the $f(p_1,p_2)$ the fact that
the S matrix can in principle be computed using Feynman rules implies that this function should not contain factors of $\pi$. 
One may remove such a contribution by adding to the ansatz \eqref{extra1} further terms based on the integral in fig.~\ref{6pt-vertex}. 
Adding such terms may also be used to repair {\it e.g.} a potential lack of factorization of the three-particle cuts of the ansatz in \eqref{Ansatz 2-loop}.

In addition to the potential cancellation described above, which mirrors the pattens of the undeformed near-flat space calculation, 
it is useful to also note the different dependence on the Jacobian factor $J$ of the integrals in figs.~\ref{2loop_rational} and \ref{QC prop} 
and fig.~\ref{2loop_integrals}.  The fact that the former integrals depend on a single external momentum implies that, unlike the latter integrals, 
their expression cannot contain any factors of $J$. This suggests that integrals having two-particle $t$-channel cuts contribute to parts 
of the two-loop S matrix that are distinct from those that receive contributions from integrals having $s$- and $u$-channel cuts.\footnote{This 
statement ignores potential $J$ factors that may exist in the integral coefficients.}
In turn this observation implies that the rational part of the one-loop dressing phase (contributing to the two-loop S matrix multiplied by 
the tree-level S matrix, cf. eq.~\eqref{SmatExpansion}) should be present in the terms written explicitly in equation \eqref{2-loop S matrix without t}. 
This is in fact the case: the last line of that equation is
\begin{align}
-\frac{i(1-\nu^2)^{3/2}}{16\pi J} (C_g+C_h)
=&{}\frac{i}{2\pi}\sqrt{1-\nu^2}\frac{(\omega_1^2-1)(\omega_2^2-1)}{\omega_2p_1-p_2\omega_1}(i\tmatrix^{(0)})
=\frac{i(1-\nu^2)^{3/2}}{8\pi} C_t \, (i\tmatrix^{(0)})\ ,
\end{align}
which indeed reproduces the contribution of the rational part of the one-loop dressing phase to the two-loop S matrix, 
cf. eq.~\eqref{SmatExpansion} and the discussion in  sec.~\ref{sec:1loopStrue}, eqs.~\eqref{1loopSfull} and \eqref{Ct1loop}.

\section{Discussion}\label{Diskussion}

In this paper we discussed in detail string theory in $\eta$-deformed AdS$_5\times$S$^5$ and we have seen 
that, for the perturbative worldsheet S matrix to be consistent with integrability and the expected PSU$_q(2|2)^2$ of the gauge-fixed 
theory the naive tree-level two-particle asymptotic states (and more generally all multi-particle asymptotic states) must 
be redefined non-locally. We have checked that the exponentiation of the redefinition required by the one-loop S matrix renders 
consistent the two-loop S matrix as well suggesting that this exponentiation may be exact to all loop orders. 
It would of course be interesting to check whether this is indeed the case.

The necessity for such a redefinition, which does not parallel the undeformed theory, 
is related to the presence of the deformation and in particular to the fact that the 
worldsheet theory contains a nontrivial bosonic Wess-Zumino term. Since the worldsheet symmetry generators have a nontrivial 
expansion around the BMN vacuum, it is possible that the redefinition we identified is necessary for the two-particle 
state to be a representation of the PSU$(2|2)_q$ symmetry of the gauge-fixed theory at the quantum level.  It would be 
interesting to explore the properties that a nontrivial NS-NS background should have for such a redefinition to be necessary.

We have also identified a set of two-loop scalar and tensor integrals that capture all the logarithmic terms in the two-loop 
S matrix. By evaluating the integrals we have observed that the same expression also captures correctly 
some rational terms, in particular the rational terms corresponding to the contribution of the one-loop dressing phase to the 
two-loop S matrix. 
Other such terms however are not; attempting to understand them we pointed out that a certain cancellation pattern between 
external line corrections and $t$-channel integrals can occur provided that one allows for the presence of integrals that have 
only three-particle cuts.
It goes without saying that a complete understanding of the rational terms of the two- and higher-loop S matrix remains an important 
open problem.

Ideally, one would expect that, to any loop order, it should be possible to write an integral representation for the S matrix of the form
\begin{align}
i\tmatrix^{(L)}&=\sum_j\; \frac{1}{S_j^{(L)}} \,C_j^{(L)} I_j^{(L)} \ ,
\end{align}
\noindent 
where $I_j^{(L)}$ are integrals with only four-point vertices, $C_j^{(L)}$ are cuts written entirely in terms of the four-point 
tree-level S matrix and $S_j^{(L)}$ are symmetry factors of the corresponding integrals. There is a certain amount of freedom 
in choosing the integrals $I_j^{(L)}$ and not all choices need to be consistent with integrability, in particular the fact that cuts 
isolating tree-level higher-point amplitudes are factorized.
Thus, apart from the integrals listed above, tree-level integrability may require inclusion of integrals with higher-point vertices as well.

With the appropriate definition of propagators, the integrals identified here can be used to construct the massive S matrix in all  
AdS$_n\times$S$^n\times$M$^{10-2n}$ spaces and, presumably, in all two-dimensional integrable theories. For $n<5$ however 
an important issue that awaits a satisfactory resolution is the contribution of massless modes. It has been suggested in 
\cite{Bianchi:2014rfa} and verified explicitly in \cite{Roiban:2014cia} through Feynman graph calculation that they do not 
contribute to the S matrix at one-loop level. It would be interesting to understand the reason behind this feature and whether
their decoupling continues at higher loops as well.

\section*{Acknowledgements:}

We would like to thank S.~Frolov, B.~Hoare and A.~Tseytlin for useful discussions and B.~Hoare, J.~Minahan and A.~Tseytlin for comments on the draft.
RR acknowledges the  hospitality of the KITP at UC Santa Barbara
and, while this work was being written up, also that of the Simons Center for Geometry and Physics. 
The work of OTE is supported by the Knut and Alice Wallenberg Foundation under grant KAW~2013.0235.
The work of RR was supported in part by the US Department of Energy under contract DE-SC0008745 
and while at KITP also by the National Science Foundation under Grant No. NSF PHY11-25915. 

\newpage

\appendix

\section{Tree-level S-matrix coefficients \label{TreeSmatrix}}

In this appendix we collect the tree-level expressions of the coefficients of the various tensor structures parametrizing the S matrix, see eq.~\eqref{cTmatr}:
\begin{align}
A^{(0)}(p_1,p_2)=&{}\frac{1-2a}{4}(\omega_2p_1-p_2\omega_1)+\frac{1}{4}\frac{(p_1-p_2)^2+\nu^2(\omega_1-\omega_2)^2}{\omega_2p_1-p_2\omega_1},\nonumber\\
B^{(0)}(p_1,p_2)=&{}-E^{(0)}(p_1,p_2)=\frac{p_1p_2+\nu^2\omega_1\omega_2}{\omega_2p_1-p_2\omega_1},\nonumber\\
C^{(0)}(p_1,p_2)=&{}F^{(0)}(p_1,p_2)=\frac{1}{2}\frac{\sqrt{(p_1-i\nu)(p_1+i\nu\omega_1)(p_2-i\nu)(p_2+i\nu\omega_2)}}{\omega_2p_1-p_2\omega_1}\nonumber\\
&\phantom{F(p_1,p_2)=}\frac{(1+\omega_1)(p_2+i\nu)-(1+\omega_2)(p_1+i\nu)}{\sqrt{(1-\nu^2)(1+\omega_1)(1+\omega_2)}},\nonumber\\
D^{(0)}(p_1,p_2)=&{}\frac{1-2a}{4}(\omega_2p_1-p_2\omega_1)-\frac{1}{4}\frac{(p_1-p_2)^2+\nu^2(\omega_1-\omega_2)^2}{\omega_2p_1-p_2\omega_1},\nonumber\\
G^{(0)}(p_1,p_2)=&{}-L^{(0)}(p_2,p_1)=\frac{1-2a}{4}(\omega_2p_1-p_2\omega_1)-\frac{1}{4}\frac{\omega_1^2-\omega_2^2}{\omega_2p_1-p_2\omega_1},\\
H^{(0)}(p_1,p_2)=&{}K^{(0)}(p_1,p_2)=\frac{1}{2}\frac{\sqrt{(p_1-i\nu)(p_1+i\nu\omega_1)(p_2-i\nu)(p_2+i\nu\omega_2)}}{\omega_2p_1-p_2\omega_1}\nonumber\\
&\phantom{K(p_1,p_2)=}\frac{(1-\nu^2)(1+\omega_1)(1+\omega_2)-(p_1+i\nu)(p_2+i\nu)}{(1-\nu^2)\sqrt{(1+\omega_1)(1+\omega_2)}},\nonumber\\
W^{(0)}_B(p_1,p_2)=&{}W^{(0)}_E(p_1,p_2)=i\nu,\nonumber\\
V^{(0)}_B(p_1,p_2)=&{}V^{(0)}_E(p_1,p_2)=0,\nonumber\\
Q^{(0)}_C(p_1,p_2)=&{}Q^{(0)}_F(p_1,p_2)=0,\nonumber\\
R^{(0)}_C(p_1,p_2)=&{}R^{(0)}_F(p_1,p_2)=0.\nonumber
\end{align}

\section{Dispersion relation, propagator and Jacobian \label{various_things}}

The deformation changes the dispersion relation which in turn affects our calculations in numerous different ways. In this 
appendix we include some of the affected quantities as well as some useful identities. The dispersion relation to leading order 
in the large $g$ expansion is:
\begin{align}
\omega={}\sqrt{\frac{1+p^2}{1-\nu^2}}
\label{disprel}
\end{align}

From this  it is not hard to show that
\begin{align}
(p+i\nu)(p-i\nu)=&{}(1-\nu^2)(\omega^2-1) \ ,\\
(p+i\nu\omega)(p-i\nu\omega)=&{}(\omega^2-1),
\end{align}
\noindent 
which are helpful for rewriting some of the off-diagonal S-matrix elements.

One of the quantities that appears often in generalized unitarity calculations comes from the normalization of wave-functions and 
from the Jacobian that arises when solving the energy-momentum conserving delta function in terms of constraints on space-like momenta. 
This quantity is modified by the deformation as follows:
\begin{align}
&\frac{1}{4\omega_1\omega_2}\delta^2(\vec{p}_1+\vec{p}_2-\vec{p}_3-\vec{p}_4)
\\
&\qquad\qquad
={}\frac{1-\nu^2}{4(\omega_2p_1-\omega_1p_2)}\left[\delta(p_1-p_3)\delta(p_2-p_4)+\delta(p_1-p_4)\delta(p_2-p_3)\right]\nonumber.
\end{align}
We shall denote the overall factor on the right-hand side by $J$:
\begin{align}
J=&{}\frac{4(\omega_2p_1-\omega_1p_2)}{1-\nu^2} \ .
\end{align}

The $\nu$-dependence of the dispersion relation \eqref{disprel} implies that the propagators are changed into:
\begin{align}
\Delta(\omega, q) = \frac{1}{\omega^2-q^2/(1-\nu^2)-1/(1-\nu^2)},
\end{align}
\noindent 
and consequently the integrals also need to be modified compared to the $\nu=0$  case. 
The simplest way to see how the deformation affects them is to rescale the space-like momenta and thus obtain a two-dimensional Lorentz 
invariant propagator with mass
\be
m=\frac{1}{\sqrt{1-\nu^2}}
\ee 
All integrals have therefore the same form as in the un-deformed theory up to rescaling of the space-like momentum. 
For convenience we define 
\be
p_{-}=\omega-\frac{p}{\sqrt{1-\nu^2}} \ .
\ee

\section{The difference of $s$- and $u$-channel one-loop integral coefficients \label{CsMinusCu}}

In this appendix we collect the differences of the matrix elements of the $C_s$ and $C_u$ one-loop integral coefficients.
\bea
\label{ABcoef}
\frac{(C_s)_{ab}^{cd}}{J}-\frac{(C_u)_{ab}^{cd}}{J}&=&{}\delta_a^c\delta_b^d((B^{(0)})^2+2C^{(0)}F^{(0)}-(W^{(0)})^2)
-\delta_a^d\delta_b^c(2C^{(0)}F^{(0)}+2(B^{(0)})^2
\\
&-&2H^{(0)}K^{(0)}
-2(W^{(0)})^2+(W^{(0)})^2\epsilon_{ab}(\epsilon_{ab}+\epsilon^{dc})+B^{(0)}W^{(0)}(\epsilon_{ab}+\epsilon^{dc})) \ ,
\nonumber\\[2pt]
\label{DEcoefs}
\frac{(C_s)_{\alpha\beta}^{\gamma\delta}}{J}-\frac{(C_u)_{\alpha\beta}^{\gamma\delta}}{J}
&=&{}\delta_\alpha^\gamma\delta_\beta^\delta((E^{(0)}){}^2+2C^{(0)}F^{(0)}-(W^{(0)})^2)
-{}\delta_\alpha^\delta\delta_\beta^\gamma(2C^{(0)}F^{(0)}+2(E^{(0)})^2
\\
&-&2H^{(0)}K^{(0)}-2(W^{(0)})^2+(W^{(0)})^2\epsilon_{\alpha\beta}(\epsilon_{\alpha\beta}+\epsilon^{\delta\gamma})+E^{(0)}W^{(0)}(\epsilon_{\alpha\beta}+\epsilon^{\delta\gamma})) \ ,
\nonumber\\[2pt]
\label{GLcoefs}
\frac{(C_s)_{a\beta}^{c\delta}}{J}-\frac{(C_u)_{a\beta}^{c\delta}}{J}&=&{}\delta_a^c\delta_\beta^\delta(H^{(0)}K^{(0)}+C^{(0)}F^{(0)}) \ ,
\\[2pt]
\frac{(C_s)_{\alpha b}^{\gamma d}}{J}-\frac{(C_u)_{\alpha b}^{\gamma d}}{J}&=&{}\delta_\alpha^\gamma\delta_b^d(H^{(0)}K^{(0)}+C^{(0)}F^{(0)}) \ ,
\\[2pt]
\frac{(C_s)_{a\beta}^{\gamma d}}{J}-\frac{(C_u)_{a\beta}^{\gamma d}}{J}&=&{}\delta_a^d\delta_\beta^\gamma H^{(0)}
\big(G^{(0)}+L^{(0)}-A^{(0)}-D^{(0)}-2B^{(0)}-2E^{(0)}
\nonumber\\
&&-W^{(0)}(\epsilon_{a1}+\epsilon_{a2}+\epsilon_{3\beta}+\epsilon_{4\beta})\big) \ ,
\\[2pt]
\label{Kcoef}
\frac{(C_s)_{\alpha b}^{c\delta}}{J}-\frac{(C_u)_{\alpha b}^{c\delta}}{J}&=&{}\delta_\alpha^\delta\delta_b^c K^{(0)}\big(G^{(0)}+L^{(0)}-A^{(0)}-D^{(0)}-2B^{(0)}-2E^{(0)}
\nonumber\\
&&-W^{(0)}(\epsilon_{\alpha3}+\epsilon_{\alpha4}+\epsilon_{1b}+\epsilon_{2b})\big) \ ,
\\[2pt]
\frac{(C_s)_{ab}^{\gamma\delta}}{J}-\frac{(C_u)_{ab}^{\gamma\delta}}{J}&=&{}\epsilon_{ab}\epsilon^{\gamma\delta}\, C^{(0)}\big(A^{(0)}+D^{(0)}-B^{(0)}-E^{(0)}-G^{(0)}-L^{(0)}-W^{(0)}(\epsilon_{ab}+\epsilon^{\delta\gamma})\big)
\\[2pt]
\label{Fcoef}
\frac{(C_s)_{ab}^{\gamma\delta}}{J}-\frac{(C_u)_{ab}^{\gamma\delta}}{J}&=&{}\epsilon_{ab}\epsilon^{\gamma\delta}\, F^{(0)}\big(A^{(0)}+D^{(0)}-B^{(0)}-E^{(0)}-G^{(0)}-L^{(0)}-W^{(0)}(\epsilon_{ab}+\epsilon^{\delta\gamma})\big) \ .
\eea

\section{One-loop S-matrix coefficients  \label{oneloopSmatrixcoefs}}

In this appendix we collect the one-loop expressions of the coefficients parametrizing the S matrix, see eq.~\eqref{cTmatr}, in terms of their tree-level values. 
\begin{align}
iA^{(1)}=&{}\frac{-i}{2\pi}\left(H^{(0)}K^{(0)}+C^{(0)}F^{(0)}\right)\mathrm{ln}\left(\frac{p_{2-}}{p_{1-}}\right)
+\frac{i\sqrt{1-\nu^2}}{2\pi}\frac{(\omega_1^2-1)(\omega_2^2-1)}{\omega_2p_1-p_2\omega_1}\nonumber\\
&-\frac{1}{2}\left[(A^{(0)})^2+H^{(0)}K^{(0)}+C^{(0)}F^{(0)}\right] \ ,\\
iB^{(1)}=&{}-\left[A^{(0)}B^{(0)}-C^{(0)}F^{(0)}+\tfrac{1}{2}(W^{(0)})^2\right] \ ,\\
iW_B^{(1)}=&{}-A^{(0)}W^{(0)}_B \ ,\\
iV^{(1)}_B=&{}\tfrac{1}{2}(W^{(0)}_B)^2 \ ,\\
iD^{(1)}=&{}\frac{-i}{2\pi}\left(H^{(0)}K^{(0)}+C^{(0)}F^{(0)}\right)\mathrm{ln}\left(\frac{p_{2-}}{p_{1-}}\right)
+\frac{i\sqrt{1-\nu^2}}{2\pi}\frac{(\omega_1^2-1)(\omega_2^2-1)}{\omega_2p_1-p_2\omega_1}\nonumber\\
&-\frac{1}{2}\left[(D^{(0)})^2+H^{(0)}K^{(0)}+C^{(0)}F^{(0)}\right] \ ,\\
iE^{(1)}=&{}-\left[D^{(0)}E^{(0)}-C^{(0)}F^{(0)}+\tfrac{1}{2}(W^{(0)})^2\right] \ ,\\
iW^{(1)}_E=&{}-D^{(0)}W^{(0)}_E \ ,\\
iV^{(1)}_E=&{}\tfrac{1}{2}(W^{(0)}_E)^2 \ ,\\
iG^{(1)}=&{}\frac{-i}{2\pi}\left(H^{(0)}K^{(0)}+C^{(0)}F^{(0)}\right)\mathrm{ln}\left(\frac{p_{2-}}{p_{1-}}\right)
+\frac{i\sqrt{1-\nu^2}}{2\pi}\frac{(\omega_1^2-1)(\omega_2^2-1)}{\omega_2p_1-p_2\omega_1}\nonumber\\
&-\frac{1}{2}\left[(G^{(0)})^2+H^{(0)}K^{(0)}\right] \ ,\\
iL^{(1)}=&{}\frac{-i}{2\pi}\left(H^{(0)}K^{(0)}+C^{(0)}F^{(0)}\right)\mathrm{ln}\left(\frac{p_{2-}}{p_{1-}}\right)
+\frac{i\sqrt{1-\nu^2}}{2\pi}\frac{(\omega_1^2-1)(\omega_2^2-1)}{\omega_2p_1-p_2\omega_1}\nonumber\\
&-\frac{1}{2}\left[(L^{(0)})^2+H^{(0)}K^{(0)}\right] \ ,\\
iC^{(1)}=&{}-\frac{1}{2}C^{(0)}\left[A^{(0)}+D^{(0)}-B^{(0)}-E^{(0)}\right] \ ,\\
iQ^{(1)}_C=&{}-\frac{1}{4}C^{(0)}\left[W^{(0)}_B+W^{(0)}_E\right] \ ,\\
iR^{(1)}_C=&{}0 \ ,\\
iF^{(1)}=&{}-\frac{1}{2}F^{(0)}\left[A^{(0)}+D^{(0)}-B^{(0)}-E^{(0)}\right] \ ,\\
iQ^{(1)}_F=&{}-\frac{1}{4}F^{(0)}\left[W^{(0)}_B+W^{(0)}_E\right] \ ,\\
iR^{(1)}_F=&{}0 \ ,\\
iH^{(1)}=&{}-\frac{1}{2}H^{(0)}\left[G^{(0)}+L^{(0)}\right] \ ,\\
iK^{(1)}=&{}-\frac{1}{2}K^{(0)}\left[G^{(0)}+L^{(0)}\right] \ .
\end{align}

\section{One- and two-loop integrals \label{integrals}}

In terms of the $p_-$ momentum defined in Appendix \ref{various_things}, $p_{-}=\omega-\frac{p}{\sqrt{1-\nu^2}}$, the $s$-, $u$- and 
$t$-channel one-loop integrals are given by 
\begin{align}
I_s=&{}\frac{1}{J}\left(-\frac{i}{\pi}\mathrm{ln}\left(\frac{p_{2-}}{p_{1-}}\right)-1\right) \ ,\\
I_u=&{}\frac{1}{J}\left(+\frac{i}{\pi}\mathrm{ln}\left(\frac{p_{2-}}{p_{1-}}\right)+0\right) \ ,\\
I_t=&{}\frac{i(1-\nu^2)^{3/2}}{4\pi} \ .
\end{align}

The ten two-loop scalar tensor integrals may be evaluated in terms of the explicit two-loop integrals of \cite{Klose:2007rz} and are given by:
\begin{align}
I_a=&{}\left(\frac{1}{J}\left(-\frac{i}{\pi}\mathrm{ln}\left(\frac{p_{2-}}{p_{1-}}\right)-1\right)\right)^2 \ ,\nonumber\\
I_d=&{}\left(\frac{1}{J}\left(+\frac{i}{\pi}\mathrm{ln}\left(\frac{p_{2-}}{p_{1-}}\right)+0\right)\right)^2 \ ,\nonumber\\
I_b=&{}-\frac{1}{4}\left(\frac{1}{J^2}\left(-\frac{i}{\pi}\mathrm{ln}\left(\frac{p_{2-}}{p_{1-}}\right)
-1\right)^2-\frac{(1-\nu^2)^3}{32\big((1-\nu^2)\omega_1\omega_2-p_1p_2-1\big)}\right) \ ,\nonumber\\
I_c=&{}-\frac{1}{4}\left(\frac{1}{J^2}\left(-\frac{i}{\pi}\mathrm{ln}\left(\frac{p_{2-}}{p_{1-}}\right)
-1\right)^2-\frac{(1-\nu^2)^3}{32\big((1-\nu^2)\omega_1\omega_2-p_1p_2-1\big)}\right) \ ,\nonumber\\
I_e=&{}-\frac{1}{4}\left(\frac{1}{J^2}\left(+\frac{i}{\pi}\mathrm{ln}\left(\frac{p_{2-}}{p_{1-}}\right)
+0\right)^2+\frac{(1-\nu^2)^3}{32\big((1-\nu^2)\omega_1\omega_2-p_1p_2+1\big)}\right) \ ,
\label{2loopints}\\
I_f=&{}-\frac{1}{4}\left(\frac{1}{J^2}\left(+\frac{i}{\pi}\mathrm{ln}\left(\frac{p_{2-}}{p_{1-}}\right)
+0\right)^2+\frac{(1-\nu^2)^3}{32\big((1-\nu^2)\omega_1\omega_2-p_1p_2+1\big)}\right) \ ,\nonumber\\
I_g=&{}\frac{1}{2}\left(\frac{i(1-\nu^2)^{3/2}}{4\pi}\right)\frac{1}{J}\left(-\frac{i}{\pi}\mathrm{ln}\left(\frac{p_{2-}}{p_{1-}}\right)
-1\right)\nonumber \ ,\\
I_h=&{}\frac{1}{2}\left(\frac{i(1-\nu^2)^{3/2}}{4\pi}\right)\frac{1}{J}\left(-\frac{i}{\pi}\mathrm{ln}\left(\frac{p_{2-}}{p_{1-}}\right)
-1\right)\nonumber \ ,\\
I_k=&{}\frac{1}{2}\left(\frac{i(1-\nu^2)^{3/2}}{4\pi}\right)\frac{1}{J}\left(+\frac{i}{\pi}\mathrm{ln}\left(\frac{p_{2-}}{p_{1-}}\right)
+0\right)\nonumber \ ,\\
I_l=&{}\frac{1}{2}\left(\frac{i(1-\nu^2)^{3/2}}{4\pi}\right)\frac{1}{J}\left(+\frac{i}{\pi}\mathrm{ln}\left(\frac{p_{2-}}{p_{1-}}\right)
+0\right)\nonumber \ .
\end{align}

For the calculations used in subsection \ref{wild} we will furthermore need the following integrals:

\begin{align}
R_a=&{}\left(\frac{i(1-\nu^2)}{4\pi}\right)^2,\nonumber\\
R_b+R_e=&{}\left(\frac{1-\nu^2}{8}\right)^2\left(\frac{1}{12}-\frac{1}{\pi^2}\right)+\frac{1}{3}\left(\frac{1-\nu^2}{8}\right)^2\frac{p_{2-}^2}{p_{1-}^2-p_{2-}^2},\nonumber\\
R_c+R_f=&{}\left(\frac{1-\nu^2}{8}\right)^2\left(\frac{1}{12}-\frac{1}{\pi^2}\right)+\frac{1}{3}\left(\frac{1-\nu^2}{8}\right)^2\frac{p_{1-}^2}{p_{1-}^2-p_{2-}^2},\nonumber\\
R_d=&{}\left(\frac{i(1-\nu^2)}{4\pi}\right)^2,\\
R'_g+R'_h=&{}3\left(\frac{1-\nu^2}{8}\right)^2\left(\frac{1}{\pi^2}-\frac{1}{12}\right),\nonumber\\
R'_k+R'_l=&{}3\left(\frac{1-\nu^2}{8}\right)^2\left(\frac{1}{\pi^2}-\frac{1}{12}\right),\nonumber\\
R_0=&{}\frac{1-\nu^2}{64}.\nonumber
\end{align}

\end{document}